\documentclass[twocolumn,secnumarabic,amssymb, nobibnotes, aps, pre ]{revtex4-1}

\usepackage{graphicx}
\usepackage{amsmath}
\usepackage{lipsum}
\usepackage{color}
\usepackage{hyperref}
\begin{document}

\title{Impact of Droplets on Inclined Flowing Liquid Films}
\author{Zhizhao Che}
\email{z.che@imperial.ac.uk}
\author{Amandine Deygas}%
\author{Omar K. Matar}%
\email{o.matar@imperial.ac.uk}
\affiliation{Department of Chemical Engineering, Imperial College London, SW7 2AZ, UK
}

\date{\today}

\begin{abstract}
The impact of droplets on an inclined falling liquid film is studied experimentally using high-speed imaging. The falling film is created on a flat substrate with controllable thicknesses and flow rates. Droplets with different sizes and speeds are used to study the impact process under various Ohnesorge and Weber numbers, and film Reynolds numbers.  A number of phenomena associated with droplet impact are identified and analysed, such as bouncing, partial coalescence, total coalescence, and splashing.  The effects of droplet size, speed, as well the film flow rate are studied culminating in the generation of an impact regime map. The analysis of the lubrication force acted on the droplet via the gas layer shows that a higher flow rate in the liquid film produces a larger lubrication force, slows down the drainage process, and increases the probability of droplet bouncing. Our results demonstrate that the flowing film has a profound effect on the droplet impact process and associated phenomena, which are markedly more complex than those accompanying impact on initially quiescent films.
\end{abstract}

\maketitle

\section{Introduction}\label{sec:sec1}
Droplet impact on solid or liquid interfaces is a ubiquitous and fascinating phenomenon in nature \cite{Martin1993DropletImpact, Yarin2005Review}, and has a wide range of applications; these include inkjet-printing, spray-painting, spray-cooling, internal combustion engines, fire suppression, deposition of solder bumps on printed circuit boards, surface-cleaning, and cell-printing.  The outcome after a droplet impacts on a surface is a fascinating problem and has received considerable attention in the literature dating back to over a century ago \cite{Worthington1908Book}.  Various impacting substrates have been employed, ranging from dry \cite{Kolinski2014PRL} and pre-wetted surfaces \cite{Cossali1997}, surfaces with physical \cite{Range1998RoughImpact} and chemical heterogeneities \cite{Chen2010wettabilityImpact, Kannan2008hydrophobicImpact}, thin \cite{Wang2000thinLiquidFilm} and deep liquid layers \cite{Adomeit2000PIV, Blanchette2009DropletCoal}, porous media \cite{Ding2011PorousJFM}, heated \cite{Bernardin1997, Tran2012PRLLeidenfrost}, sublimating \cite{Antonini2013SublimationImpact}, and electrically-conducting \cite{Deng2010Electric}, 
elastic membranes \cite{Pepper2008ElaticMembrane}, and granular materials \cite{Marston2010OnPowder}.  Non-Newtonian \cite{Bergeron2000NaturePolymerAdditive, Smith2010PolynerAdditivesPRL}, surfactant-laden \cite{Gatne2009Surfactant}, and liquid-metallic droplets \cite{Aziz2000LiquidMetal, Dykhuizen1994LiquidMetal} have also been studied. Different outcomes of impact have been observed, such as bouncing, coalescence, partial coalescence, splashing. The transitions among different regimes have been investigated \cite{Cossali1997, Rioboo2003, Wang2000thinLiquidFilm}.

Depending upon the choice of impacting surfaces and droplet physical properties, the impact dynamics can be extremely complex. Thus, analysing the accompanying phenomena during the impact process, and understanding the physics behind the phenomena are of critical importance.  Generally, the impact dynamics of droplets is dominated by the interplay between surface tension, inertia, and viscous forces; this may be modified by the presence of non-Newtonian and/or physico-chemical and/or thermal effects.  The relative importance of these forces can be quantified using dimensionless numbers, such as the Ohnesorge and Weber numbers. These dimensionless numbers are used widely in the investigation of droplet impact, e.g., in defining the regimes of impact \cite{Pan2007} and in developing empirical correlations \cite{Scheller1995, Cossali1997}.

Although droplet impact phenomena have been investigated for more than a century, it is the recent development of high-speed imaging that allows researchers to unveil the fast evolving dynamic features of the phenomena \cite{Thoroddsen2008HighSpeedPhotography}.  Many phenomena occurring during impact remain poorly understood, and continue to attract interest.  These phenomena include the collapse of the gas layer beneath the droplet during the impact \cite{deRuiter2012GasLayer, Driscoll2011GasLayerInterference, Tran2013GasLayerJFM}, the entrapment of air bubbles when a droplet impacts on a liquid surface \cite{Bouwhuis2012BubbleTrapPRL, Hasan1990JFM}, and the ejection of liquid sheets during the impact and its subsequent breakup into secondary droplets \cite{Thoraval2012LiquidSheetPRL, Thoroddsen2012LiquidSheetJFM}. Bird et al.\ \cite{bird2009inclinedSplashing} found that during the impact of droplets on angled or moving surfaces, the asymmetry leads to an azimuthal variation of the ejected rim, and the tangential component of impact can act to enhance or suppress a splash. \v{S}ikalo et al.\ \cite{Sikalo2005Inclined} studied the impact of droplets on inclined dry and wetted surfaces, and found the bouncing on dry smooth or wetted surfaces, and no bouncing on rough surfaces. Gilet and Bush \cite{Bush2012WetInclinedImpact} studied the bouncing of droplets on an inclined substrate with a thin layer of high-viscosity fluid, and analysed the contact time and the conversion of energy. The liquid layer with high viscosity was used to ensure a smooth impact surface which facilitates droplet bouncing. Despite the large number of studies on droplet impact, the impact of droplets on flowing liquid films, to the best of our knowledge, has not been studied. This is surprising since it is a common phenomenon occurring in many two-phase flows, such as annular flows \cite{Hewitt1978book}, spray-cooling \cite{Kim2007SprayCoolingReview}, and spray-painting. The interaction between the droplet and the flowing liquid film could produce unique dynamic features distinct from the impact phenomena on quiescent liquids. This will be the subject of the present paper.

A falling liquid film is used as the flowing liquid film for droplet impact in this study.  Falling films are common in nature, important to many industrial applications, and are characterised by rich nonlinear dynamics \cite{Chang1994FallingFilmReview, Chang2002FFbook, CrasterMatar2009RMP, Kalliadasis2011FFbook}.  The pioneering study of Kapitza was performed half a century ago \cite{Kapitza1948FallingFilm}.  Since then, the flow in falling liquid films has been studied extensively through experimental measurements \cite{Adomeit2000PIV, Zhou2009FFthickness}, low-dimensional modelling \cite{Scheid2006FallingFilm3DJFM, Shkadov1967Model}, and full numerical simulations \cite{Gao2003FFVOF}.  The effects of many factors that influence the flow in falling films have been investigated, such as the effects of thermocapillarity \cite{Frank2006FFThermocapillary}, electric fields \cite{Jayaratne1964, Tseluiko2006FFElec}, and surfactants \cite{Ji1994FFsurfactant, Strobel1969FFsurfactant}.  Different processes that may be involved in falling films have also been studied, such as heat transfer \cite{Scheid2008FallingFilm3DTemperatureEPL}, mass transfer \cite{Yang1992FFadsorption}, chemical reactions \cite{Dabir1996FFreaction}, and phase change \cite{Palen1994FFevaporation}.  Even though the flow of falling liquid films has been widely studied in the literature, it still remains an important research topic and attracts the attention of researchers of various disciplines. In this study, we use high-speed imaging to examine the impact of droplets of controlled diameter, angle of incidence and speed, on isothermal falling films in order to explore how the film dynamics influence the impact phenomena. In addition to the Ohnesorge and Weber numbers, this system is also parameterised by the film Reynolds number, which is based on the film mean thickness and velocity. The results of our work show that the film dynamics have a profound effect on the impact process, giving rise to phenomena that are not observed in the quiescent film case.

The rest of this paper is organised as follows. The experimental system is described in Section \ref{sec:sec2}, including the falling film rig, the droplet generation unit, and the high-speed imaging system.  The experimental results are presented in Section \ref{sec:sec3}.  A regime map of droplet impact is created, and the effects of the Weber, Ohnesorge, and Reynolds numbers are studied by varying the impact speeds, the droplet sizes, and the flow rates of the falling film. A simple model based on lubrication theory is used to explain the bouncing phenomenon.   Finally, concluding remarks are provided in Section \ref{sec:sec4}.

\section{Experimental method}\label{sec:sec2}
\subsection{Falling liquid film}

\begin{figure}
  \centering
  \includegraphics[scale=0.6]{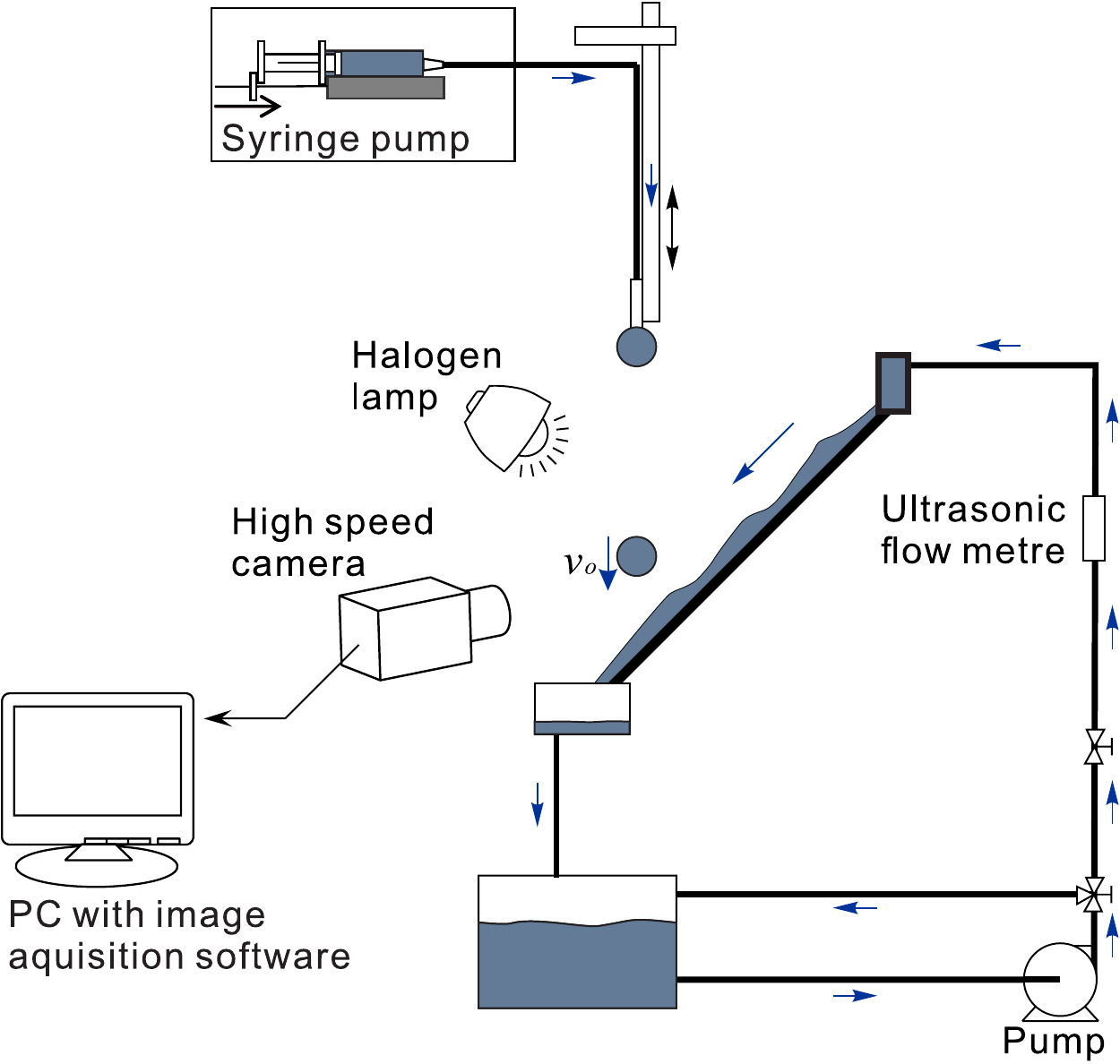}
  \caption{Schematic diagram of the experimental setup for droplet impact on falling liquid films which comprises a falling film rig, a droplet-generation unit, and a high-speed imaging system. }\label{fig:fig01}
\end{figure}

The experimental setup used to study droplet impact on flowing liquid films is shown in Figure \ref{fig:fig01}.  The falling film rig comprised a rigid and impermeable plate 41 cm long and 28 cm wide made of a titanium foil with a 50 $\mu$m thickness inclined at 45$^\circ$ to the horizontal.  The liquid used was deionised water of density $10^3$ kg/m$^3$, viscosity $10^{-3}$ Pa$\cdot$s, and surface tension 72.8 mN/m, and the flow rate was controlled by a valve and a variac transformer for the pump.  An ultrasonic flow metre (Flownetix 100 series, Flownetix, UK) was used to measure the flow rate of the falling film, which was in the range 0.9$\pm$0.1--10.7$\pm$0.3 l/min. The lower limit of the flow rate of the liquid film is the minimum flow rate that can ensure the whole substrate is wetted. The mean film thicknesses $h$ and the mean flow speed $\bar{u}$ were estimated using the Nusselt solution \cite{Kalliadasis2011FFbook} to be in the range 0.29--0.65 mm, and 0.19--0.98 m/s, respectively.  The Reynolds numbers of the falling films $\text{Re}\equiv \rho q/(w \mu) $ were therefore in the range 54--636, where $\rho$ and $\mu$ are the density and the viscosity of the liquid, $q$ and $w$ are the flow rate and the width of the falling liquid film, respectively. Since laminar-turbulence transition in falling films happens at $\text{Re}>1000$ \cite{Chang1994FallingFilmReview}, the flow in the falling liquid film studied here is laminar.

\subsection{Droplet generation unit}
Droplets of uniform size were generated from the tip of blunt syringe needles.  Deionised (DI) water was pumped into the needle using a syringe pump (Braintree Scientific Inc., UK) at constant flow rates.  The droplets broke up at the tip of the syringe needle in the dripping mode, fell under their own weights, and impacted obliquely onto the inclined liquid film.  The dripping mode and the gravity-driven acceleration of the droplets ensure that the process is highly reproducible.  To study the effect of droplet size on the impact process, the droplet size was varied via a change in the diameter of the needle. Three droplet diameters were studied, 3.3, 4.0, and 5.2 mm with standard deviation less than 0.1 mm, while the  corresponding Ohnesorge number,$\text{Oh}\equiv \mu /\sqrt{\rho \gamma d}$, were 0.0021, 0.0019, and 0.0016, respectively.  Here, $\rho $ is the density of the liquid, $\gamma$ is the surface tension, $\mu$ is the dynamic viscosity, and $d$ is the droplet diameter before impact.  To study the effect of the droplet speed on the impact process, the impact speeds of the droplets were varied by changing the vertical distance of the needle towards the falling liquid film from 1 to 50 cm.  The corresponding speeds of the droplet $v_0$ were in the range 0.30$\pm$0.02--2.99$\pm$0.09 m/s, and the corresponding Weber numbers, $\text{We}\equiv \rho dv_{0}^{2}/\gamma $, were in the range of 4--630 for different droplet sizes.  The impact point on the liquid film is 34 cm from the inlet of the falling film, which is larger than 500 times of the film thickness of the Nusselt solution \cite{Kalliadasis2011FFbook} at the highest flow rate studied, and allows the development of the film flow on the substrate. The details of the flow in the falling film will be discussed in Subsection \ref{sec:sec3_1}.

\subsection{high-speed imaging system}
The droplet impact process was recorded using a high-speed camera (Olympus i-SPEED 3) at a frame rate of 5000 frames per second. A macro lens (Sigma EX 105mm) was used to capture the details during the impact. Two halogen lamps (500W) were used for illumination. To minimize the heating effect of the halogen lamps, the lamps were kept on only when the camera was recording. Side-view images were captured at a resolution of 804$\times$396 pixels (about 40 $\mu$m/pixel) with 10$^\circ$ deviation from the plane of the film, while top view images were capture at a resolution of 528$\times$600 pixels perpendicular to the film (about 40 $\mu$m/pixel). The high-speed images were analysed using customised Matlab routines.  The speed and the size of the droplets were measured from the high-speed images.

\section{Results and discussion}\label{sec:sec3}
In this section, we present a discussion of our results starting with a brief description of the wavy falling film dynamics in the absence of droplet impact; this provides an indication of the expected state of the interface prior to droplet impact at various film $\text{Re}$. A detailed investigation of the wavy flow in the liquid film is beyond the scope of this study. We then describe the main phenomena associated with the impact process, before presenting a flow regime diagram as a function of system parameters. The results of a simple model based on lubrication theory are also discussed, and used to explain the trends associated with a subset of the phenomena observed.

\subsection{Wavy flow in liquid films}\label{sec:sec3_1}
\begin{figure}
  \centering
  \includegraphics[width=\columnwidth]{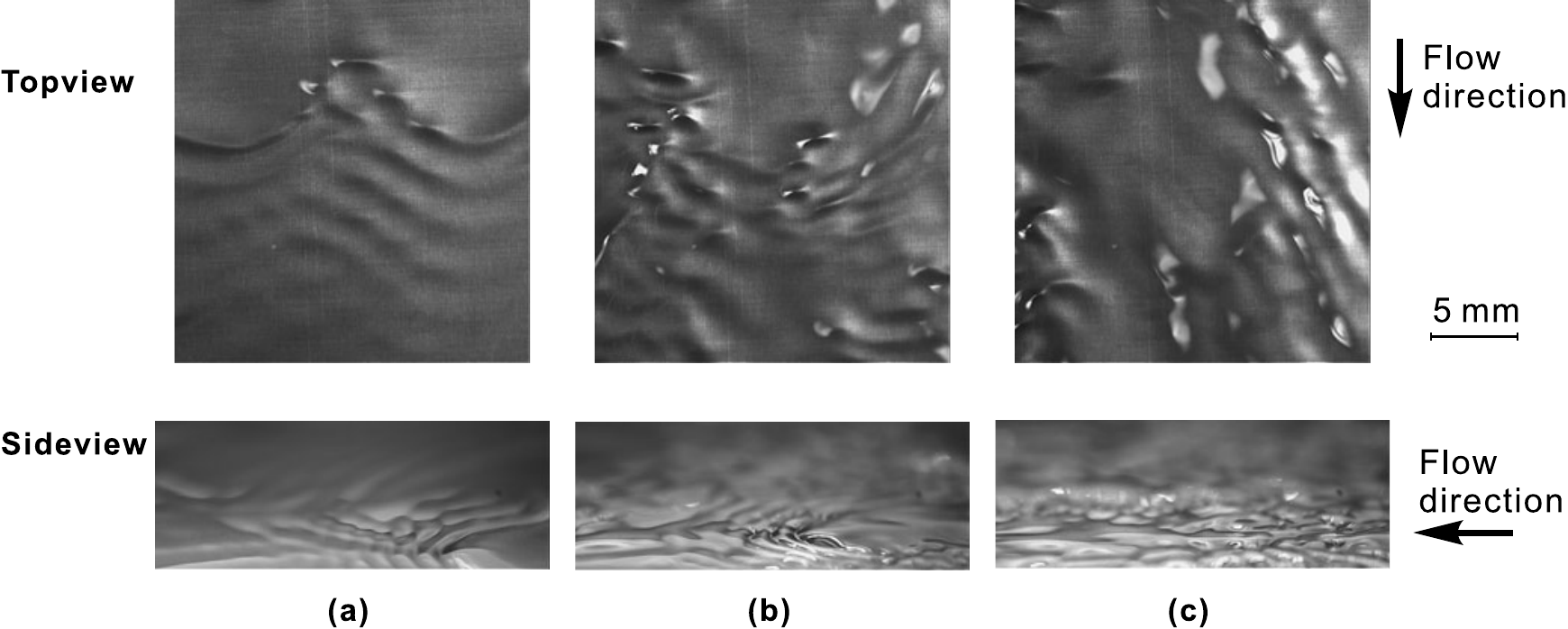}
  \caption{Typical images of the wavy flow in the falling liquid films at different Reynolds numbers: (a) $\text{Re}= 54$, (b) $\text{Re}=245$, and (c) $\text{Re}=636$. The flow rates are 0.9, 4.1, and 10.7 l/min, respectively.}\label{fig:fig02}
\end{figure}

\begin{figure}
  \centering
  \includegraphics[width=\columnwidth]{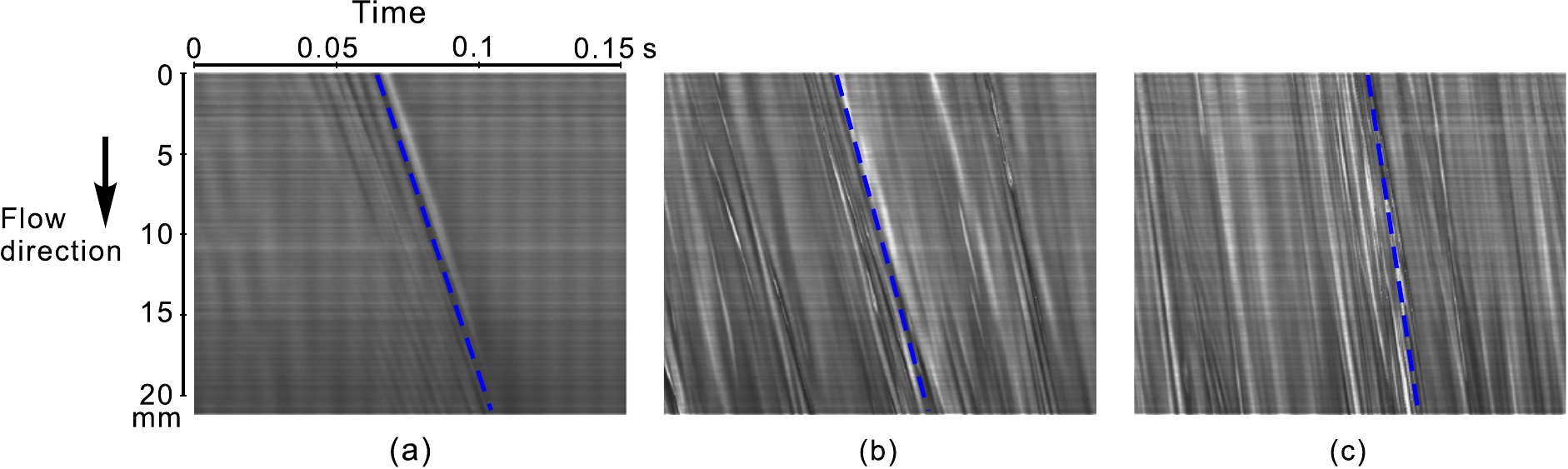}
  \caption{Time-space plots showing the evolution of the wavy flow in the falling liquid films at different Reynolds numbers: (a) $\text{Re}= 54$, (b) $\text{Re}=245$, and (c) $\text{Re}=636$. The flow rates are 0.9, 4.1, and 10.7, l/min respectively. The speeds of the waves can be estimated from the gradient of the dashed lines of the waves. }\label{fig:fig03}
\end{figure}

The wavy flow in the falling liquid films was analysed from the top-view and side-view images near the point of the droplet impact, as shown in Figure \ref{fig:fig02}. At low $\text{Re}$ values, the liquid films are characterised by interfaces that exhibit mild deformations with occasional solitary waves. These waves are formed with coherent precursor waves, as shown in Figure \ref{fig:fig02}a.  As $\text{Re}$ increases, the waves become more frequent and less coherent. The evolution of the waves can be seen from the time-space plot of a line along the flow direction, which is obtained from the top-view images, as shown in Figure \ref{fig:fig03}.
In a time-space plot, the (spatial) wave length is the vertical distance of consecutive waves, and the (temporal) frequency of the waves is the reciprocal of the horizontal distance of consecutive waves. The amplitude of the waves could not be measured in this experiment quantitatively due to the limitation of the experimental facility. The variation of the film thickness, however, is represented by the light intensity. Previous studies of falling liquid films \cite{Chang1994FallingFilmReview,Chang2002FFbook,CrasterMatar2009RMP,Kalliadasis2011FFbook} have shown that the amplitude of the waves is usually smaller or comparable to the mean film thickness, and the wave lengths are much larger than the mean film thickness. The surface curvature of the liquid film is very small as compared to the droplet curvature due to the small aspect ratio of the liquid film (thickness vs film length).
At low $\text{Re}$, the solitary wave in Figure \ref{fig:fig03}a travels downstream at a constant speed. With increasing $\text{Re}$, the waves interact with adjacent waves as they propagate downstream, as shown in Figure \ref{fig:fig03}b-c. The speed of the waves can be estimated from the gradient of the line of the waves in the time-space plot, which shows that the wave speed increases with $\text{Re}$.

In the following subsections, the impact processes on falling films at different conditions of $\text{We}$, $\text{Re}$, and $\text{Oh}$ numbers are analysed. Various combinations of the parameters result in different phenomena as observed in the experiments: bouncing, partial coalescence, total coalescence, and splashing, as will be detailed below. Because the waves on the liquid films are formed naturally and stochastically, the drop may impact on different parts of the waves. Therefore, the outcome of the impact may vary, and the results are presented in a probability manner.

\subsection{Bouncing}\label{sec:sec3_2}
\begin{figure}
  \centering
  \includegraphics[scale=0.5]{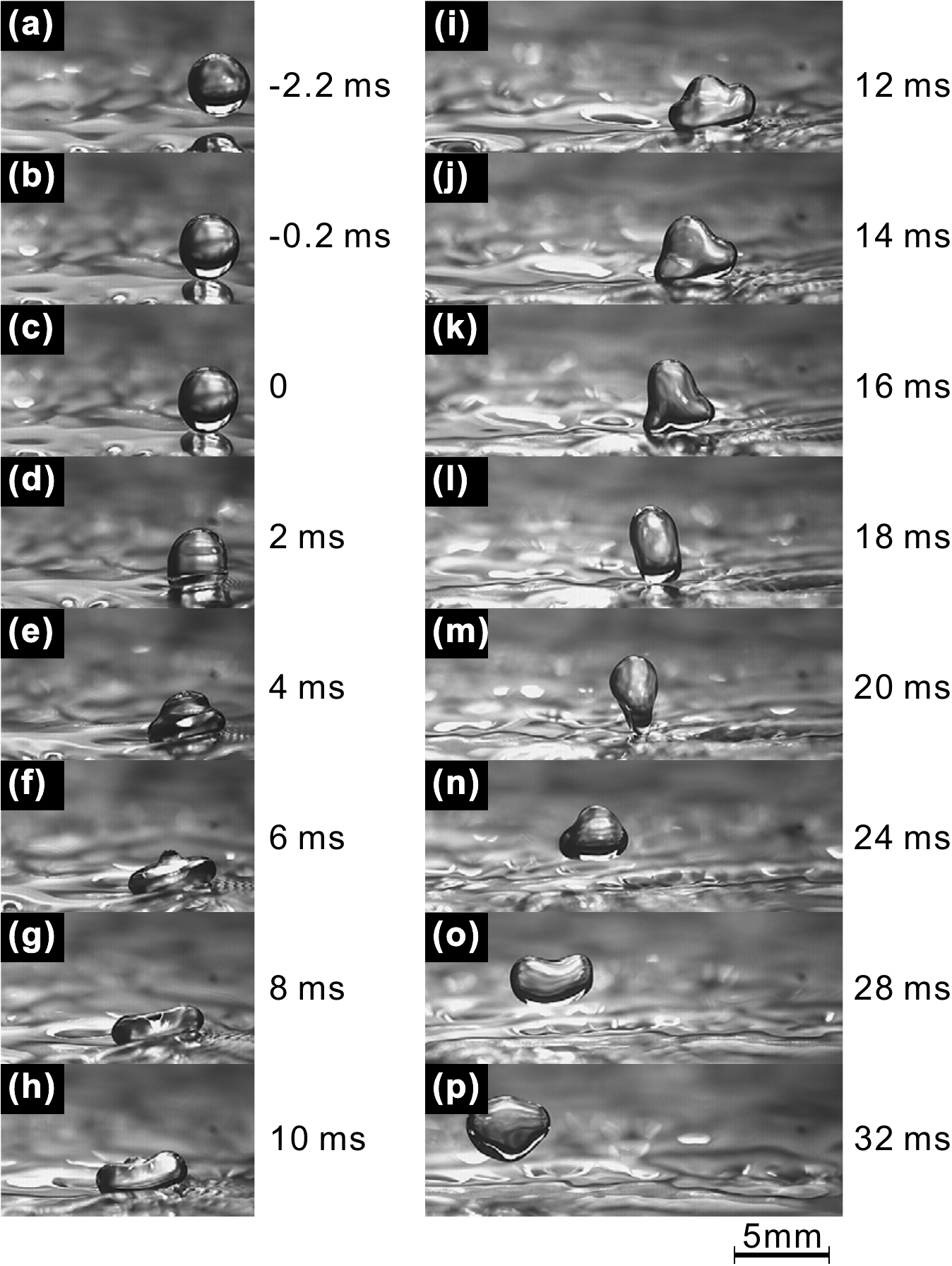}
  \caption{`Bouncing' of a droplet on a falling liquid film. The dimensionless parameters are $\text{Oh}=0.0021$, $\text{We}=4$, and $\text{Re}=636$. The droplet speed is 0.3 m/s, the flow rate of the falling film is 10.7 l/min, and the droplet diameter is 3.3 mm. See Ref.\ \cite{RefSM}, Movie 1.
  }\label{fig:fig04}
\end{figure}

`Bouncing' refers to situations wherein a droplet approaching the falling film, does not coalesce with it, but instead bounces back, as shown in Figure \ref{fig:fig04}, which was generated for $\text{Oh}=0.0021$, $\text{We}=4$, and $\text{Re}=636$. The coalescence of a droplet with a liquid film involves three stages: (i) droplet approach towards the film; (ii) the thinning of the thin, intervening gas layer between the droplet and the film; and (iii) the rupture of the gas layer.  The first stage is driven by the momentum of the droplet gained from gravity, the second stage is resisted by the viscous force in the thin gas layer, and the third stage is dominated by capillary force. As the gas layer becomes thinner, the resistance becomes larger, according to lubrication theory \cite{Davis1989Lubrication, Deen1998TransportPhenomena}.  Therefore, the gas layer could potentially stabilise the droplet against coalescence. Rupture of the gas layer occurs when its thickness decreases below a certain level so that van der Waals forces come into operation to destabilise the layer \cite{Hahn1985CoalescenceVanDerWaals}.

Prior to impact, the droplet is almost spherical (see Figure \ref{fig:fig04}a), as the capillary force minimizes the surface area and, in turn, the surface energy. Upon impact, the droplet progressively adopts a pyramidal (see Figure \ref{fig:fig04}e) and then a pancake shape (see Figure \ref{fig:fig04}g).
The latter increases the contact area of the droplet with the film and further retards the thinning of the gas layer, according to lubrication theory \cite{Davis1989Lubrication, Deen1998TransportPhenomena}.
The surface energy of the droplet, $E_\gamma$, can be estimated through image analysis by assuming a surface of revolution, ${{E}_{\gamma }}\equiv S\gamma $, where $S$ is the surface area of the droplet. The result shows that from the undistorted droplet shape (see Figure \ref{fig:fig04} c) to the maximum deformation (see Figure 4g), the surface energy increases from approximately $2.5\times10^{-6}$ J to $3.4\times10^{-6}$ J. Meanwhile, the kinetic energy before the impact, ${{E}_{v}}\equiv \frac{1}{2}m{{v}_{0}}^{2}$, and the loss of potential energy, ${{E}_{g}}\equiv mg\Delta z$, are approximately $8\times10^{-7}$ J and $5\times10^{-7}$ J, respectively. Here, $m$ is the mass of the droplet, and $\Delta z$ is the vertical distance that the droplet travels from the instant of apparent contact to the instant of maximum deformation.  Therefore, the relative conversion of energy from kinetic and potential energy to surface energy during the process can be calculated from
\begin{equation}\label{eq:eq_Econversion}
    \varepsilon \equiv \frac{\Delta {{E}_{\gamma }}}{{{E}_{v}}+{{E}_{g}}},
\end{equation}
which is estimated to be about 70\%. Such high ratio of energy conversion has also been found for droplet bouncing on horizontal and inclined wet surfaces \cite{Bush2012WetInclinedImpact}.

The Weber number is a key parameter that affects the thinning of the gas layer prior to possible rupture: lower Weber numbers (lower impact speeds) lead to longer gas layer drainage times, and, therefore, an increased probability of droplet bouncing. The momentum of the falling film, on the other hand, for a fixed inclination angle, increases with the film Reynolds number (the liquid flow rate).  Therefore, droplets have a greater chance of bouncing on liquid films with higher Reynolds numbers.  This is consistent with the phenomenon of droplet levitation at the hydraulic jump point of a liquid film produced from a vertical jet impinging on a horizontal surface \cite{Sreenivas1999LevitationJFM}. The details will be explained in detail in Section \ref{sec:sec3_7a}.

\subsection{Partial coalescence}\label{sec:sec3_3}

\begin{figure}
  \centering
  \includegraphics[scale=0.5]{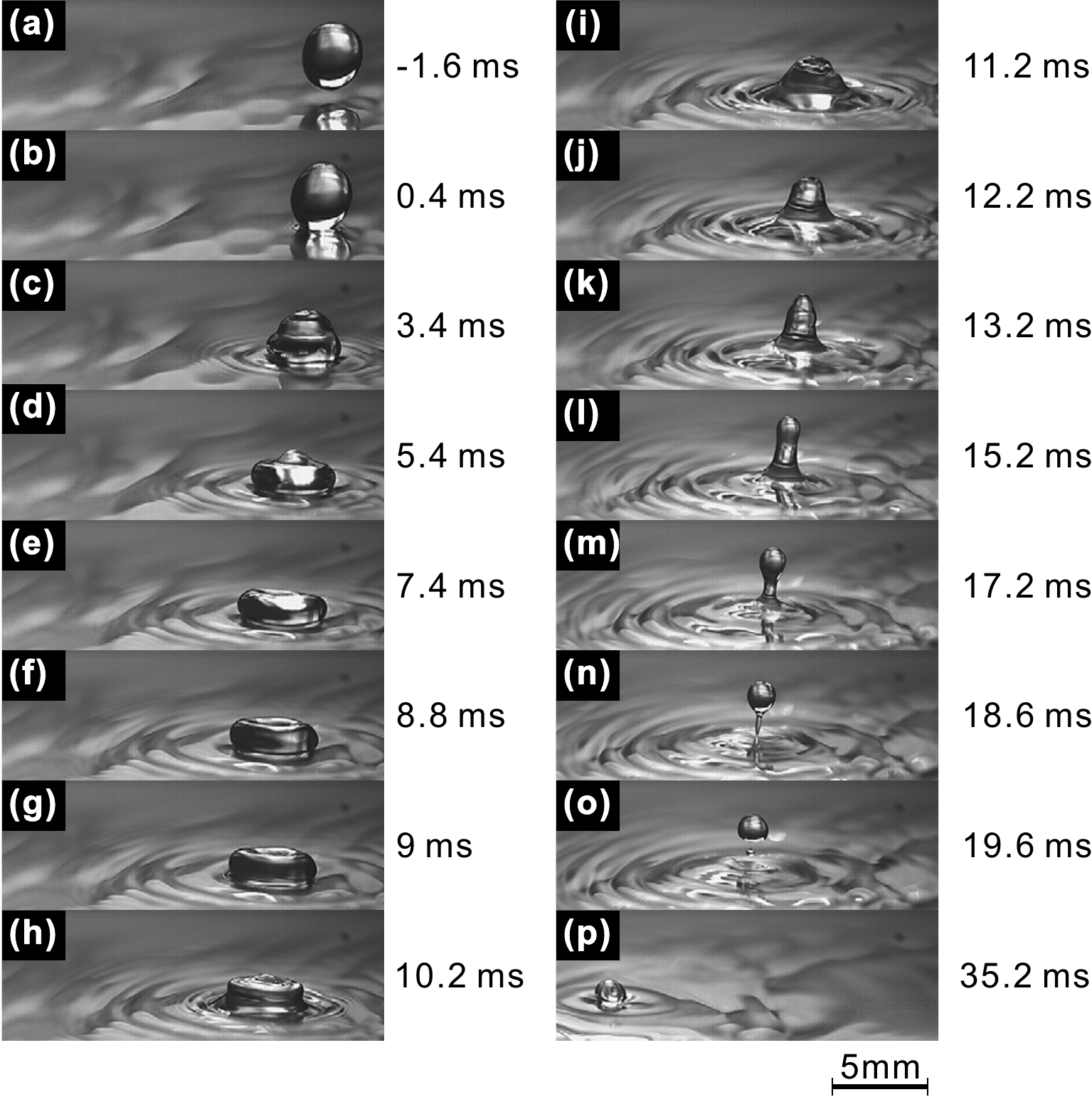}
  \caption{`Partial coalescence' of a droplet on a falling liquid film.  The dimensionless parameters are $\text{Oh}=0.0021$, $\text{We}=4$, and $\text{Re}=54$. The droplet speed is 0.3 m/s, the flow rate of the falling film is 0.9 l/min, and the droplet diameter is 3.3 mm. See Ref.\ \cite{RefSM}, Movie 2.
  }\label{fig:fig05}
\end{figure}

For a range of system parameters, the droplets were observed to undergo `partial coalescence' with the falling film. This corresponds to situations wherein, upon impacting the film, a fraction of the droplet coalesces with the film, but immediately after impact, a small daughter droplet or a cascade of daughter droplets \cite{Blanchette2006PartialCoalescence, Gilet2007PartialCoalescence} are created; an example of partial coalescence is shown in Figure \ref{fig:fig05}.  It occurs as a consequence of small Weber numbers and small film Reynolds numbers.

\begin{figure}
  \centering
  \includegraphics[scale=0.6]{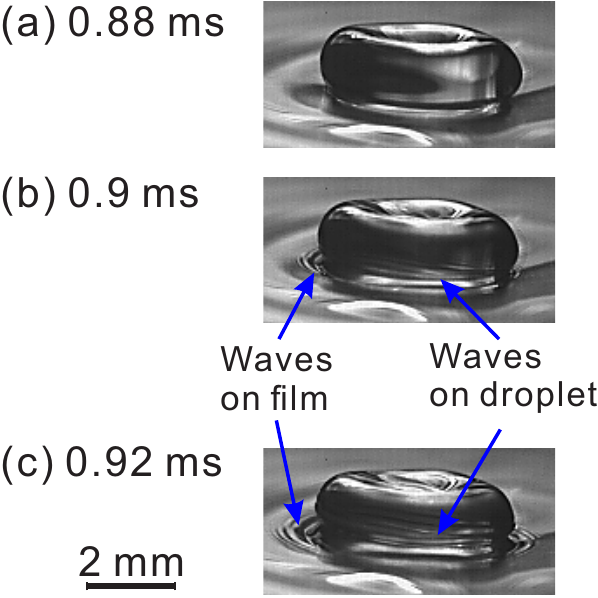}
  \caption{Formation and propagation of capillary waves immediately after the collapse of the gas layer between the droplet and the liquid film. The dimensionless parameters are $\text{Oh}=0.0021$, $\text{We}=4$, and $\text{Re}=54$. The droplet speed is 0.3 m/s, the flow rate of the falling film is 0.9 l/min, and the droplet diameter is 3.3 mm.
  }\label{fig:fig06}
\end{figure}

The rupture of the gas layer starts with the formation of a liquid bridge connecting the droplet and the liquid film. Due to the large curvature at the liquid bridge and the corresponding large capillary force, the liquid bridge immediately grows, and produces capillary waves on the surface of the droplet and on the liquid film (see Figure \ref{fig:fig05}g and Figure \ref{fig:fig06} for the formation and the propagation of the capillary waves).  The waves on the droplet then propagate rapidly towards the apex of the droplets (see Figure \ref{fig:fig06}c).  Immediately after the collapse of the gas layer, the liquid at the bottom of the droplet is significantly affected by the capillary waves and is pulled towards the liquid film.  However, the liquid at the droplet apex is less affected, and there is no sufficient time to catch up with that near the bottom of the droplet. Therefore, the droplet becomes a long thin liquid thread (see Figure \ref{fig:fig05}l), which eventually breaks up into two parts (see Figure \ref{fig:fig05}o) due to the inward momentum of the collapsing neck \cite{Blanchette2006PartialCoalescence}. The lower part mixes with the liquid in the falling film, while the upper part gradually falls on the liquid film due to gravity as a daughter droplet.  The daughter droplet tends to be more stable against coalescence because of its smaller speed than the primary droplet before partial coalescence, as shown in Figure \ref{fig:fig05}p.

The surface energy after pinch-off of the daughter droplet (see Figure 5n) can be estimated through image analysis to be $E_{\gamma} \sim 6 \times 10^{-7}$ J. It is much smaller than the initial surface energy before the impact (see Figure 5a), $E_{\gamma} \sim 2.5 \times 10^{-6}$ J. This indicates significant energy loss due  to viscous dissipation and the formation of
capillary waves during the process.

\subsection{Total coalescence}\label{sec:sec3_4}
\begin{figure}
  \centering
  \includegraphics[width=\columnwidth]{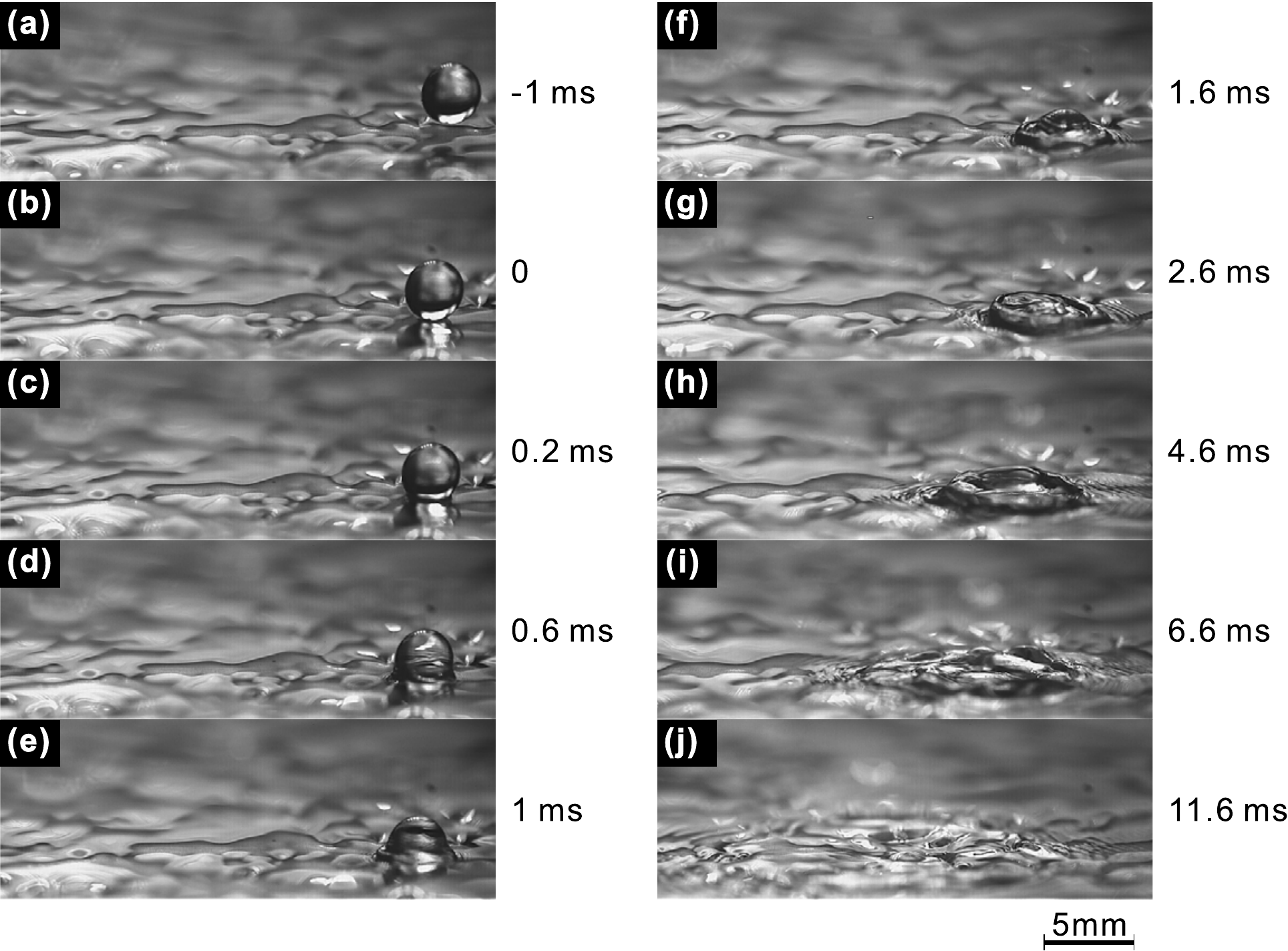}
  \caption{`Total coalescence' of a droplet on a falling film.  The dimensionless parameters are $\text{Oh}=0.0021$, $\text{We}=76$, and $\text{Re}=636$. The droplet speed is 1.3 m/s, the flow rate of the falling film is 10.7 l/min, and the droplet diameter is 3.3 mm. See Ref.\ \cite{RefSM}, Movie 3.
  }\label{fig:fig07}
\end{figure}

In contrast to the situations considered above, `total coalescence' refers to complete coalescence of the droplet with the falling film upon impact without forming smaller, secondary droplets; an example of total coalescence is shown in Figure \ref{fig:fig07} generated for $\text{Oh}=0.0021$, $\text{We}=76$, and $\text{Re}=636$. If the Weber number is above a certain level, the upper part of the droplet, follows the lower part of the droplet and coalesces with the liquid film (see Figure \ref{fig:fig07}f), without breaking up into a daughter droplet, as in the case of partial coalescence. In addition, the high-momentum liquid in the droplet pushes away the liquid in the flowing film and produces a crater (see Figure \ref{fig:fig07}g).  The waves produced by this event propagate away from the crater and dissipate gradually (see Figure \ref{fig:fig07}h).

\subsection{Splashing}\label{sec:sec3_5}
\begin{figure}
  \centering
  \includegraphics[width=\columnwidth]{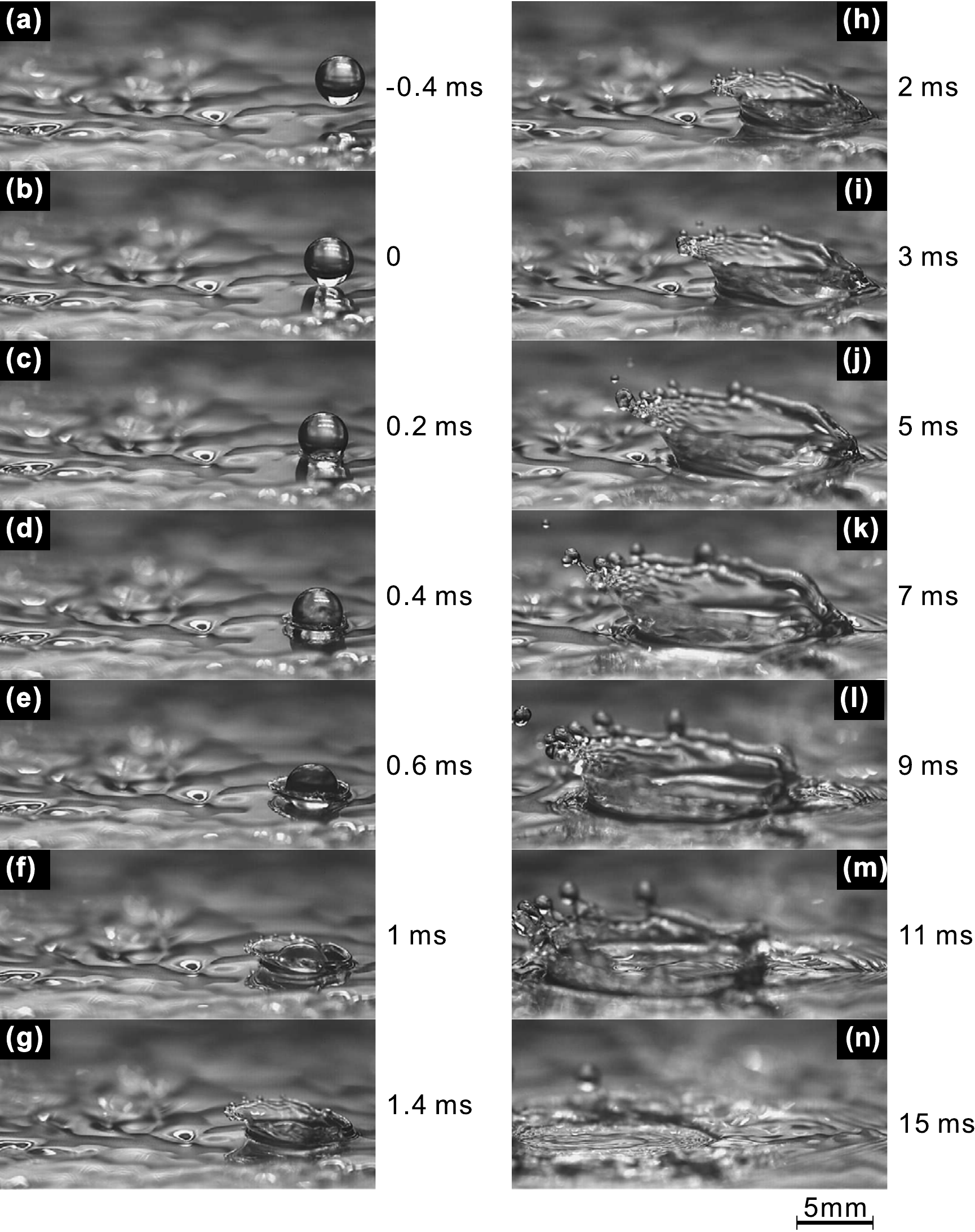}
  \caption{`Splashing' during droplet impact on a falling film.  The dimensionless parameters are $\text{Oh}=0.0021$, $\text{We}=316$, and $\text{Re}=636$.  The droplet speed is 2.67 m/s, the flow rate of the falling film is 10.7 l/min, and the droplet diameter is 3.3 mm. See Ref.\ \cite{RefSM}, Movie 4.
  }\label{fig:fig08}
\end{figure}

If the momentum of the droplet is sufficiently large, characterised by high $\text{We}$, the droplet drives the liquid in the film to form a `crown', which then disintegrates into many droplets, as shown in Figure \ref{fig:fig08}. This `splashing' phenomenon occurs because the gas layer separating the impacting droplet and the falling film undergoes rapid rupture, and capillary forces are not sufficiently strong so as to balance the large force per unit area at high Weber numbers. Thus, following impact, the liquid undergoes severe deformation leading to the formation of a cylindrical sheet of liquid around the droplet (see Figure \ref{fig:fig08}e), which then assumes a crown shape (see Figure \ref{fig:fig08}h).  The crown rim then undergoes an Rayleigh-Plateau instability \cite{Zhang2010CrownRPInstability} that results in the creation and subsequent ejection of small droplets (see Figure \ref{fig:fig08}k). The crown formation and the ejection of small droplets for the impact of droplets on quiescent liquids has been well studied in the literature \cite{Martin1993DropletImpact, Wang2000thinLiquidFilm, Yarin2005Review, Thoraval2012LiquidSheetPRL, Thoroddsen2012LiquidSheetJFM}. Being different from impact on quiescent liquids, the wavy interface of the liquid film results in a more irregular crown structure and different sizes of the small droplets. The crown is also not symmetric due to the oblique impact angle \cite{bird2009inclinedSplashing}: the rim on the downstream side is much higher than that on the upstream side of the crown, which leads to the ejection of more droplets on the downstream side. The details will be discussed in Subsection \ref{sec:sec3_8}.

\subsection{Regime map}\label{sec:sec3_6}
\begin{figure}
  \centering
  \includegraphics[width=0.9\columnwidth]{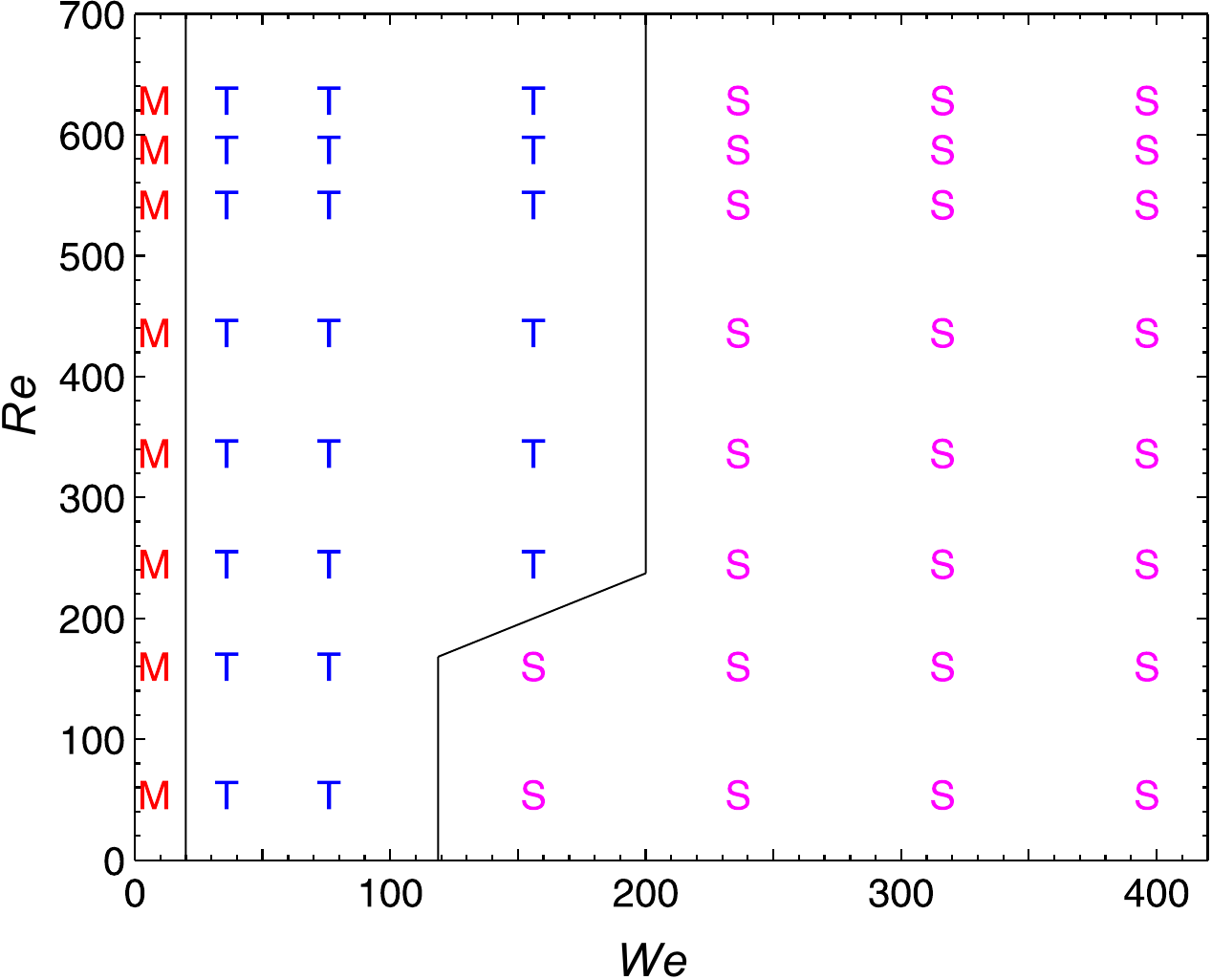}
  \caption{Regime map for droplet impact on falling liquid films.  Here, `S', `T', and `M' designate `splashing', `total coalescence', and multiple phenomena as described in Figure \ref{fig:fig10}.  The Ohnesorge number is fixed at $\text{Oh}=0.0021$.  The droplet speed is in the range of 0.30--2.99 m/s, the flow rate of the falling film is in the range of 0.9--10.7 l/min, and the droplet diameter is 3.3 mm.  The lines are used only to guide readers' eyes.
  }\label{fig:fig09}
\end{figure}

\begin{figure}
  \centering
  \includegraphics[scale=0.46]{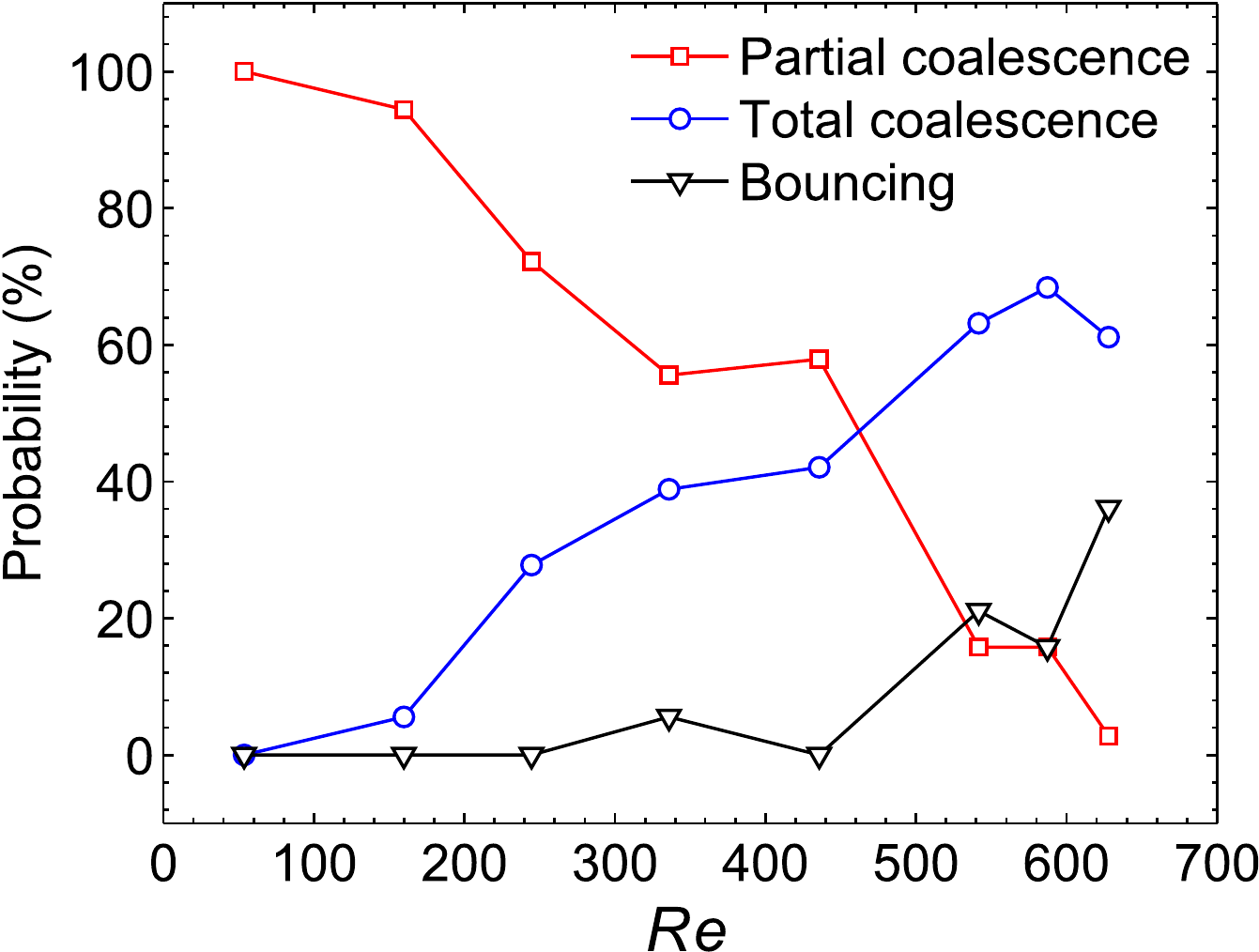}
  \caption{Probability of occurrence of different phenomena as a function of $\text{Re}$ with $\text{Oh}=0.0021$ and $\text{We}=4$.  The droplet speed is 0.30 m/s, the flow rate of the falling film is in the range of 0.9--10.7 l/min, and the droplet diameter is 3.3 mm.
  }\label{fig:fig10}
\end{figure}

A regime map for droplet impact on falling liquid films is created by varying the flow rate of the falling film and the impact speed of the droplets; this is presented in Figure \ref{fig:fig09} as a $\text{Re}$ vs $\text{We}$ plot which shows the regime transitions as these parameters are varied for $\text{Oh}=0.0021$. For a given $\text{Re}$, large $\text{We}$ values lead to splashing, while moderate $\text{We}$ values lead to total coalescence.  At the lowest $\text{We}$ studied, multiple phenomena are observed, namely bouncing, partial coalescence, and total coalescence, and their probability of occurrence in repeated experiments are plotted in Figure \ref{fig:fig10}.

As shown in Figure \ref{fig:fig10}, partial coalescence always occurs for the lowest $\text{Re}$ studied.  As $\text{Re}$ increases, the probability of partial coalescence decreases, while that of total coalescence rises. This is because a thick liquid film facilitates the flow of liquid in the direction perpendicular to the film and allows the liquid near the top region of the droplet to catch up with that near the bottom region, thereby reducing the probability of formation of secondary droplets following impact. The probability of bouncing also increases with the film Reynolds number.  This is consistent with the phenomenon of droplet levitation at the hydraulic jump point of a liquid film produced from a vertical jet impinging on a horizontal surface \cite{Sreenivas1999LevitationJFM}, which corresponds to a large $\text{Re}$ and negligible $\text{We}$ in Figure \ref{fig:fig09}.

\subsection{Lubrication gas layer and force analysis}\label{sec:sec3_7a}
To explain the phenomena presented above, the gas layer between the droplet and the liquid film and the forces involved are analysed. Prior to the rupture of the gas layer between the droplet and the liquid film, the gas layer drains gradually. The drainage of the gas layer is controlled by the competition between the inertia of the approaching droplet and the lubrication force in the gas layer. The former serves as the driving force and is indicated by the normal component of the droplet speed to the falling film, while the latter serves as the resistance and is affected by the flow speed in the film. Thus, droplet bouncing is expected only at low normal speeds of impact.
Even though the thickness of the gas layer is too thin to be determined from the observation in Figure \ref{fig:fig04}, measurements of the film thickness for droplet impact on solid surfaces in the literature \cite{VeenDrainageInterferometry2012, deRuiter2012GasLayer} have shown that the thickness of the gas layer can be down to several micrometres. To estimate the lubrication force acted on the droplet by the gas layer during the impact process, two complementary methods are presented. The first method is through an analysis of the flow in the gas layer, the liquid film, and the droplet, while the second method relies on the analysis of the droplet deformation shown on the high-speed images.

\subsubsection{Estimation from the flow in the gas layer, the liquid film, and the droplet}

\begin{figure}
  \centering
  \includegraphics[scale=0.9]{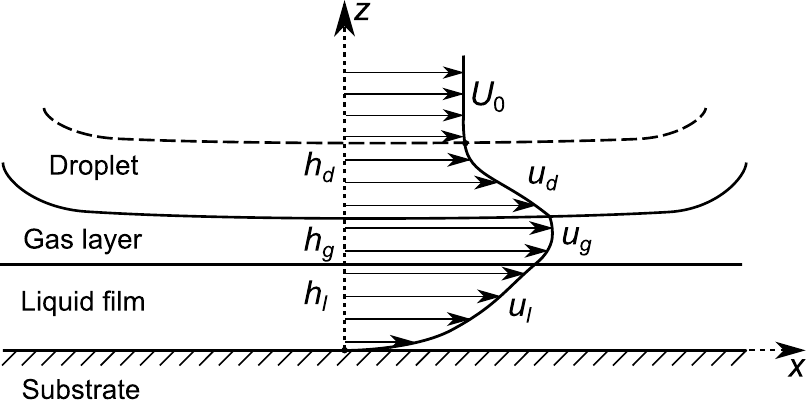}
  \caption{Schematic diagram of the flow in the gas layer and the liquid film beneath the droplet during the impact on flowing liquid film.
  }\label{fig:figA1}
\end{figure}

\begin{figure}
  \centering
  \includegraphics[scale=0.65]{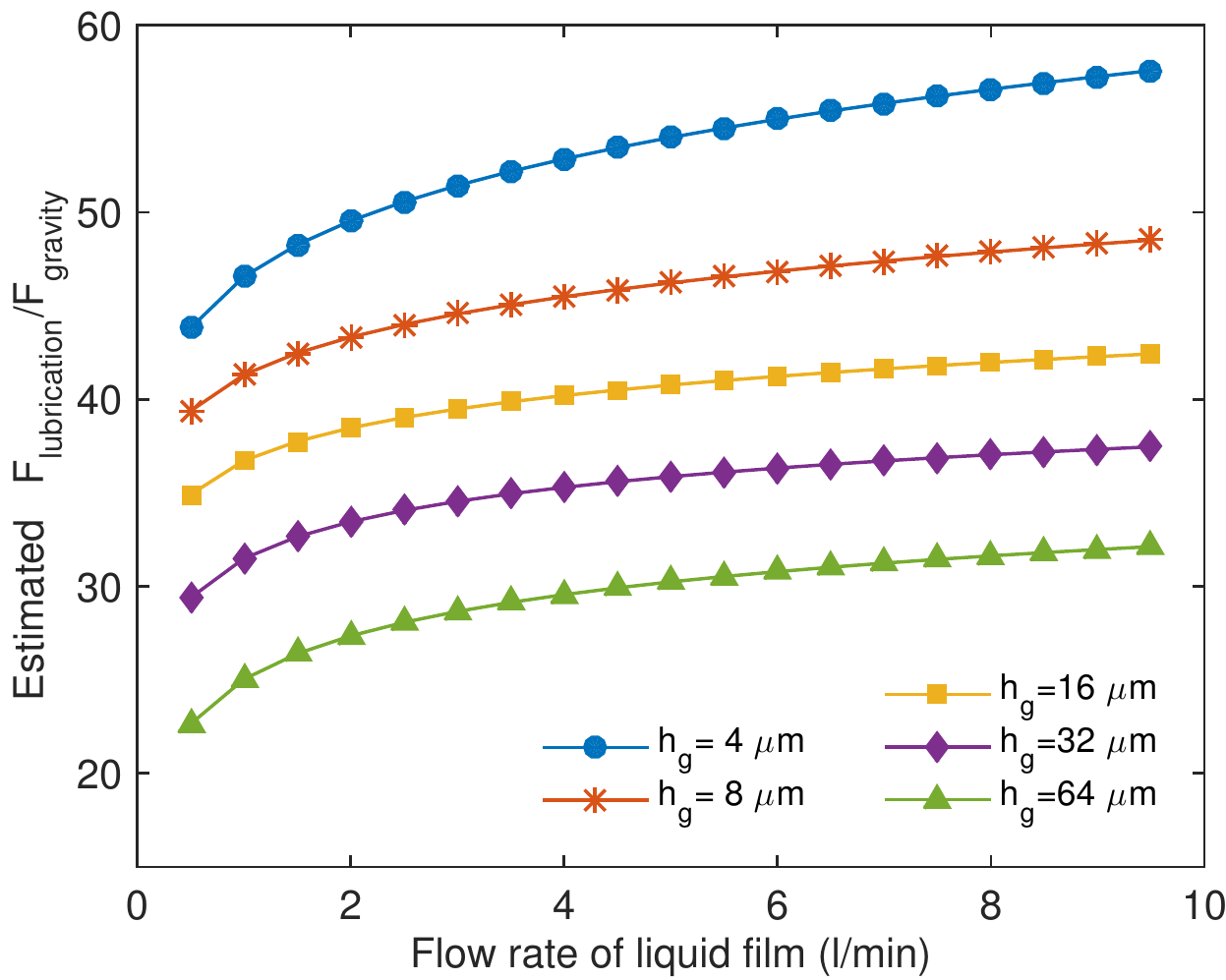}
  \caption{Estimation of the lubrication force normalised by the gravitational force normal to the film, ${{F}_{\text{gravity}}}=\frac{1}{6}\pi {{d}^{3}}\rho g\cos \left( \theta  \right)$, as a function of the flow rate of the liquid film at different thickness of the gas layer.
  }\label{fig:figA2}
\end{figure}

Due to the small thickness of the gas layer and the liquid film relative to the drainage length of the gas layer in the tangential direction, the flow in the gas layer and in the liquid film can be approximated using lubrication theory. The approximation is also applied in a penetration depth $h_d$ in the droplet, and the velocity gradient vanishes and the velocity approaches the droplet tangential speed at the penetration depth \cite{Li1996PenitrationDepth, Yeo2001233}, as shown in Figure \ref{fig:figA1}. We also neglect deformations of the film-gas interface in the present, approximate analysis:
\begin{equation}\label{eq:eqA1}
    {{\mu }_{g}}\frac{{{\partial }^{2}}{{u}_{g}}}{\partial {{z}^{2}}}=\frac{\partial {{p}_{g}}}{\partial x},
\end{equation}
\begin{equation}\label{eq:eqA2}
    {{\mu }_{l}}\frac{{{\partial }^{2}}{{u}_{l}}}{\partial {{z}^{2}}}=\frac{\partial {{p}_{l}}}{\partial x},
\end{equation}
\begin{equation}\label{eq:eqA3}
    {{\mu }_{l}}\frac{{{\partial }^{2}}{{u}_{d}}}{\partial {{z}^{2}}}=\frac{\partial {{p}_{d}}}{\partial x},
\end{equation}
where the subscript $g$ indicates the gas layer, and $l$ the liquid film, and $d$ the droplet. Then the velocity profiles along the $z$ direction are parabolic:
\begin{equation}\label{eq:eqA4}
    {{u}_{g}}={{A}_{g}}{{z}^{2}}+{{B}_{g}}z+{{C}_{g}},
\end{equation}
\begin{equation}\label{eq:eqA5}
    {{u}_{l}}={{A}_{l}}{{z}^{2}}+{{B}_{l}}z+{{C}_{l}},
\end{equation}
\begin{equation}\label{eq:eqA6}
    {{u}_{d}}={{A}_{d}}{{z}^{2}}+{{B}_{d}}z+{{C}_{d}},
\end{equation}
where the coefficients ${{A}_{g}}$, ${{B}_{g}}$, ${{C}_{g}}$, ${{A}_{l}}$, ${{B}_{l}}$, ${{C}_{l}}$, ${{A}_{d}}$, ${{B}_{d}}$, ${{C}_{d}}$ can be obtained by satisfying the following boundary conditions and closure equations. A no-slip boundary condition is imposed at the substrate, and continuity of velocity is imposed at the gas/liquid-film interface and at the gas/droplet interface,
\begin{equation}\label{eq:eqA7}
    {{\left. {{u}_{l}} \right|}_{z=0}}=0,
\end{equation}
\begin{equation}\label{eq:eqA8}
    {{\left. {{u}_{g}} \right|}_{z={{h}_{l}}}}={{\left. {{u}_{l}} \right|}_{z={{h}_{l}}}},
\end{equation}
\begin{equation}\label{eq:eqA9}
    {{\left. {{u}_{g}} \right|}_{z={{h}_{0}}}}={{\left. {{u}_{d}} \right|}_{z={{h}_{0}}}},
\end{equation}
where ${{h}_{0}}\equiv {{h}_{l}}+{{h}_{g}}$. The matching of shear stress at the gas/liquid-film interface and at the gas/droplet interface yields
\begin{equation}\label{eq:eqA10}
    {{\left. {{\mu }_{g}}\frac{\partial {{u}_{g}}}{\partial z} \right|}_{z={{h}_{l}}}}={{\mu }_{l}}{{\left. \frac{\partial {{u}_{l}}}{\partial z} \right|}_{z={{h}_{l}}}},
\end{equation}
\begin{equation}\label{eq:eqA11}
    {{\left. {{\mu }_{g}}\frac{\partial {{u}_{g}}}{\partial z} \right|}_{z={{h}_{0}}}}={{\mu }_{l}}{{\left. \frac{\partial {{u}_{d}}}{\partial z} \right|}_{z={{h}_{0}}}}.
\end{equation}
The shear stress at the penetration depth of the droplet vanishes, and the speed approaches the tangential speed of the droplet with respect to the substrate, ${{U}_{0}}$:
\begin{equation}\label{eq:eqA12}
    {{\left. {{u}_{d}} \right|}_{z={{h}_{2}}}}={{U}_{0}},
\end{equation}
\begin{equation}\label{eq:eqA13}
    {{\mu }_{l}}{{\left. \frac{\partial {{u}_{d}}}{\partial z} \right|}_{z={{h}_{2}}}}=0,
\end{equation}
where ${{h}_{2}}\equiv {{h}_{l}}+{{h}_{g}}+{{h}_{d}}$. The flow rate of the liquid film is given by
\begin{equation}\label{eq:eqA14}
    w\int_{0}^{{{h}_{l}}}{{{u}_{l}}dz}=q.
\end{equation}
The pressure gradients in the liquid film and in the gas layer are assumed to be the same because
film deformations are neglected, and normal viscous stresses correspond to high-order terms within lubrication theory; thus, Eqs.\ (\ref{eq:eqA1}) and (\ref{eq:eqA2}) yield
\begin{equation}\label{eq:eqA15}
    {{\mu }_{g}}\frac{{{\partial }^{2}}{{u}_{g}}}{\partial {{z}^{2}}}={{\mu }_{l}}\frac{{{\partial }^{2}}{{u}_{l}}}{\partial {{z}^{2}}}.
\end{equation}
Therefore, the coefficients in Eqs.\ (\ref{eq:eqA4})--(\ref{eq:eqA5}) can be obtained:
\begin{widetext}
\begin{equation}\label{eq:eqA16}
    \begin{aligned}
     &{{A}_{g}}=-\frac{3{{\mu }_{l}}{{D}_{2}}}{{{h}_{l}}^{2}{{\mu }_{g}}{{D}_{1}}}, \quad
     &{{B}_{g}}=\frac{2{{\mu }_{l}}{{D}_{3}}}{{{h}_{l}}^{2}{{\mu }_{g}}{{D}_{1}}}, \quad
     &{{C}_{g}}=-\frac{\left( {{\mu }_{g}}-{{\mu }_{l}} \right)D_4}{{{h}_{l}}{{\mu }_{g}}{{D}_{1}}}, \quad
     &{{A}_{l}}=-\frac{3{{D}_{2}}}{{{h}_{l}}^{2}{{D}_{1}}}, \quad
     &{{B}_{l}}=\frac{2{{D}_{3}}}{{{h}_{l}}^{2}{{D}_{1}}}, \quad \\
     &{{C}_{l}}=0, \quad
     &{{A}_{d}}=\frac{D_5}{{{h}_{l}}^{2}\left( {{h}_{0}}-{{h}_{2}} \right){{D}_{1}}}, \quad
     &{{B}_{d}}=-2{{A}_{d}}{{h}_{2}}, \quad
     &{{C}_{d}}={{A}_{d}}h_{2}^{2}+{{U}_{0}},\quad
\end{aligned}
\end{equation}
where
\begin{equation*}\label{eq:eqA17}
\begin{aligned}
    & {{D}_{1}}=3{{h}_{0}}^{2}{{\mu }_{g}}-3{{h}_{0}}^{2}{{\mu }_{l}}-{{h}_{l}}^{2}{{\mu }_{g}}+{{h}_{l}}^{2}{{\mu }_{l}}-3{{h}_{0}}{{h}_{2}}{{\mu }_{g}}-{{h}_{0}}{{h}_{l}}{{\mu }_{g}}+{{h}_{2}}{{h}_{l}}{{\mu }_{g}}+2{{h}_{0}}{{h}_{l}}{{\mu }_{l}} \\
    & {{D}_{2}}=\left( {{h}_{0}}{{\mu }_{g}}-{{h}_{2}}{{\mu }_{g}}-2{{h}_{0}}{{\mu }_{l}}-2{{h}_{l}}{{\mu }_{g}}+2{{h}_{l}}{{\mu }_{l}} \right){q}/{w}\;+{{U}_{0}}{{h}_{l}}^{2}{{\mu }_{g}} \\
    & {{D}_{3}}=\left( 3{{h}_{0}}^{2}{{\mu }_{g}}-3{{h}_{0}}^{2}{{\mu }_{l}}-3{{h}_{l}}^{2}{{\mu }_{g}}+3{{h}_{l}}^{2}{{\mu }_{l}}-3{{h}_{0}}{{h}_{2}}{{\mu }_{g}} \right){q}/{w}\;+{{U}_{0}}{{h}_{l}}^{3}{{\mu }_{g}} \\
    & D_4= \left( 6{{h}_{0}}^{2}{{\mu }_{l}}-6{{h}_{0}}^{2}{{\mu }_{g}}+6{{h}_{0}}{{h}_{2}}{{\mu }_{g}}+3{{h}_{0}}{{h}_{l}}{{\mu }_{g}}-3{{h}_{2}}{{h}_{l}}{{\mu }_{g}}-6{{h}_{0}}{{h}_{l}}{{\mu }_{l}} \right){q}/{w}\;+{{h}_{l}}^{3}{{\mu }_{g}}{{U}_{0}} \\
    & D_5=\left( 3{{h}_{0}}^{2}{{\mu }_{l}}-3{{h}_{l}}^{2}{{\mu }_{g}}+3{{h}_{l}}^{2}{{\mu }_{l}}+6{{h}_{0}}{{h}_{l}}{{\mu }_{g}}-6{{h}_{0}}{{h}_{l}}{{\mu }_{l}} \right){q}/{w}\;+\left( {{h}_{l}}^{3}{{\mu }_{g}}-3{{h}_{0}}{{h}_{l}}^{2}{{\mu }_{g}} \right){{U}_{0}}
\end{aligned}
\end{equation*}
\end{widetext}
Although the exact value of pressure could not be known due to the complex structure at the rim of the gas layer, the pressure difference between the gas layer and the ambient can be estimated from the pressure gradient,
\begin{equation}\label{eq:eqA18}
{{p}_{g}}-{{p}_{a}}\sim -\frac{d}{2}\frac{\partial {{p}_{g}}}{\partial x}=-\frac{d}{2}{{\mu }_{g}}\frac{{{\partial }^{2}}{{u}_{g}}}{\partial {{z}^{2}}}=-{{\mu }_{g}}{{A}_{g}}d,
\end{equation}
where we have made use of Eqs.\ (\ref{eq:eqA1}), (\ref{eq:eqA4}), (\ref{eq:eqA15}), and (\ref{eq:eqA16}). Therefore, the lubrication force can be estimated as
\begin{equation}\label{eq:eqA19}
{{F}_{\text{lubrication}}}=\int\limits_{A}{\left( {{p}_{g}}-{{p}_{a}} \right)dA}\sim -\frac{\pi }{4}{{d}^{3}}{{\mu }_{g}}{{A}_{g}}.
\end{equation}
The liquid film thickness ${{h}_{l}}$ can be estimated from the Nusselt solution for the falling liquid film \cite{Kalliadasis2011FFbook},
\begin{equation}\label{eq:eqA20}
{{h}_{l}}={{\left( \frac{3\mu q}{w\rho g\cos \theta } \right)}^{1/3}}.
\end{equation}
The penetration depth in the droplet is assumed to be $h_d=4h_g$ as in Ref.\ \cite{Yeo2001233}. Therefore, during the drainage of the gas layer, the magnitude of the lubrication force ${{F}_{\text{lubrication}}}$ can be estimated at different thicknesses of the gas layer, as shown in Figure \ref{fig:figA2}.
The lubrication force (i.e., the resistant force) being large than the normal component of the gravitational force (i.e., the driving force) is a minimum requirement of droplet bouncing. Therefore, the gravitational force of the droplet in the direction normal to the falling film, ${{F}_{\text{gravity}}}=\frac{1}{6}\pi {{d}^{3}}\rho g\cos \left( \theta  \right)$, is used as a reference to scale the estimated lubrication force, where $\theta$ is the inclination angle of the falling film to the horizontal. Figure \ref{fig:figA2} shows that as the gas layer becomes thinner, the lubrication force becomes larger. It also shows that the lubrication force increases with increasing the film flow rate. Since a large lubrication force indicates a large resistance of the droplet, a slow drainage of the gas layer, and a high chance of droplet bouncing, the result in Figure \ref{fig:figA2} is consistent with the experimental observation shown in Figure \ref{fig:fig10}.

\subsubsection{Estimation from the droplet deformation in high-speed images}
\begin{figure}
  \centering
  \includegraphics[scale=0.7]{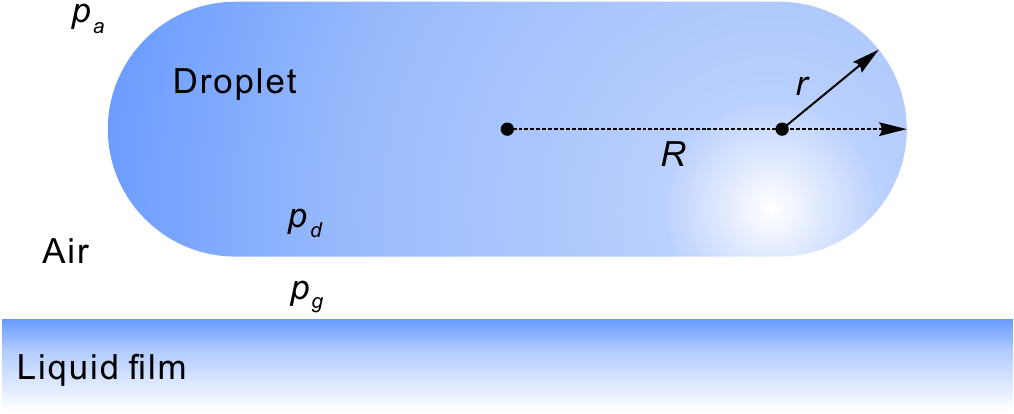}
  \caption{Schematic diagram of the droplet deformation and the pressure.
  }\label{fig:fig10a}
\end{figure}

The force applied on the droplet by the falling liquid film via the gas layer can also be estimated from the droplet deformation in the high-speed images.  When the droplet reaches the maximum deformation, it has a pancake shape, as shown in Figure \ref{fig:fig04}g. The pressure in the droplet can be estimated through the surface curvature of the droplet,
\begin{equation}\label{eq:eqL2}
{{p}_{d}}\sim \gamma \left( \frac{1}{R}+\frac{1}{r} \right)+{{p}_{a}}
\end{equation}
where the $R$ and $r$ are two principle radii of the curved droplet interface measured on the side of the non-spherical droplet, as shown Figure \ref{fig:fig10a}. Since the base of the droplet is almost flat and does not deform rapidly during the later stage of the drainage of the gas layer, the pressure in the droplet and in the gas layer can be regarded to be in equilibrium, and the pressure in the gas layer, ${{p}_{g}}$, can be approximated by the pressure in the droplet, ${{p}_{d}}$,
\begin{equation}\label{eq:eqL3}
{{p}_{g}}\sim \gamma \left( \frac{1}{R}+\frac{1}{r} \right)+{{p}_{a}}
\end{equation}
The lubrication force applied on the droplet by the liquid film via the gas layer can be estimated by multiplying the pressure in the gas layer and the base area of the droplet, ${{S}_{b}}\sim \pi {{\left( R-r \right)}^{2}}$. Therefore, the lubrication force is
\begin{equation}\label{eq:eqL4}
{{F}_{\text{lubrication}}}\sim \gamma \left( \frac{1}{R}+\frac{1}{r} \right)\pi {{\left( R-r \right)}^{2}}
\end{equation}
where the effect of the ambient pressure ${{p}_{a}}$ in Eq.\ (\ref{eq:eqL3}) is cancelled by the ambient pressure on the upper side of the droplet.
For the bouncing event shown in Figure \ref{fig:fig04}, image analysis shows that the ratio of the lubrication force to the gravitational force ${{{F}_{\text{lubrication}}}}/{{{F}_{\text{gravity}}}}\sim 10$. Therefore, the lubrication force by the gas layer on the droplet can be much larger than the gravitational force of the droplet, and can result in bouncing.

At impact, both the gas-droplet and gas-film interfaces undergo deformation \cite{Tran2013GasLayerJFM} whose severity depends on the system parameters.  An estimate of the liquid film deformation is obtained by comparing inertial, capillary, and viscous forces. The inertial force tends to deform the interface, while the capillary and viscous forces tend to resist this deformation. Their relative importance can be quantified by their ratios represented by the Weber number and the Reynolds number for film deformation, respectively given by:
\begin{equation}\label{eq:wedeform}
  \text{We}_{\text{deformation}}=\frac{\rho \,{{\left( {{v}_{0}}\cos \theta  \right)}^{2}}\,h}{\gamma }
\end{equation}
\begin{equation}\label{eq:redeform}
  \text{Re}_{\text{deformation}}=\frac{\rho \,\left( {{v}_{0}}\cos \theta  \right)\,h}{\mu }
\end{equation}
The characteristic velocity is the normal velocity of droplet towards the plate, ${{v}_{0}}\cos \theta $, while the characteristic length is the film thickness $h$ as calculated from the Nusselt solution \cite{Kalliadasis2011FFbook}.  At an impact speed of 0.30 m/s, for instance, ${\text{We}_{\text{deformation}}}=0.18$, and ${\text{Re}_{\text{deformation}}}=62$, which indicate that the surface tension is very important in resisting the deformation of the film interface.  Equations (\ref{eq:wedeform}--\ref{eq:redeform}) also indicate that thicker films which correspond to high film $\text{Re}$ are easy to deform. In addition, the relative importance between the viscous and capillary forces is quantified by
\begin{equation}\label{eq:cadeform}
  {\text{Ca}_{\text{deformation}}}=\frac{{\text{We}_{\text{deformation}}}}{{\text{Re}_{\text{deformation}}}}=\frac{\mu {{v}_{0}}\cos \theta }{\gamma }
\end{equation}
which is in the range of 0.003--0.03 for different impact speeds (0.30--2.99 m/s), indicating that the capillary force is much stronger than the viscous force.

During impact, at larger $\text{Re}$ values of the liquid films, the average liquid films are thicker and more deformable since the restoring capillary force becomes progressively weaker. The deformation of the gas/liquid-film interface delays the drainage of the gas layer, preventing its rupture. In addition, the force exerted by the film on the droplet at high film $\text{Re}$, which is transmitted through the gas layer, pushes the droplet away from the impact region. These factors also could increase the probability of droplet bouncing.

\begin{figure}
  \centering
  \includegraphics[scale=0.4]{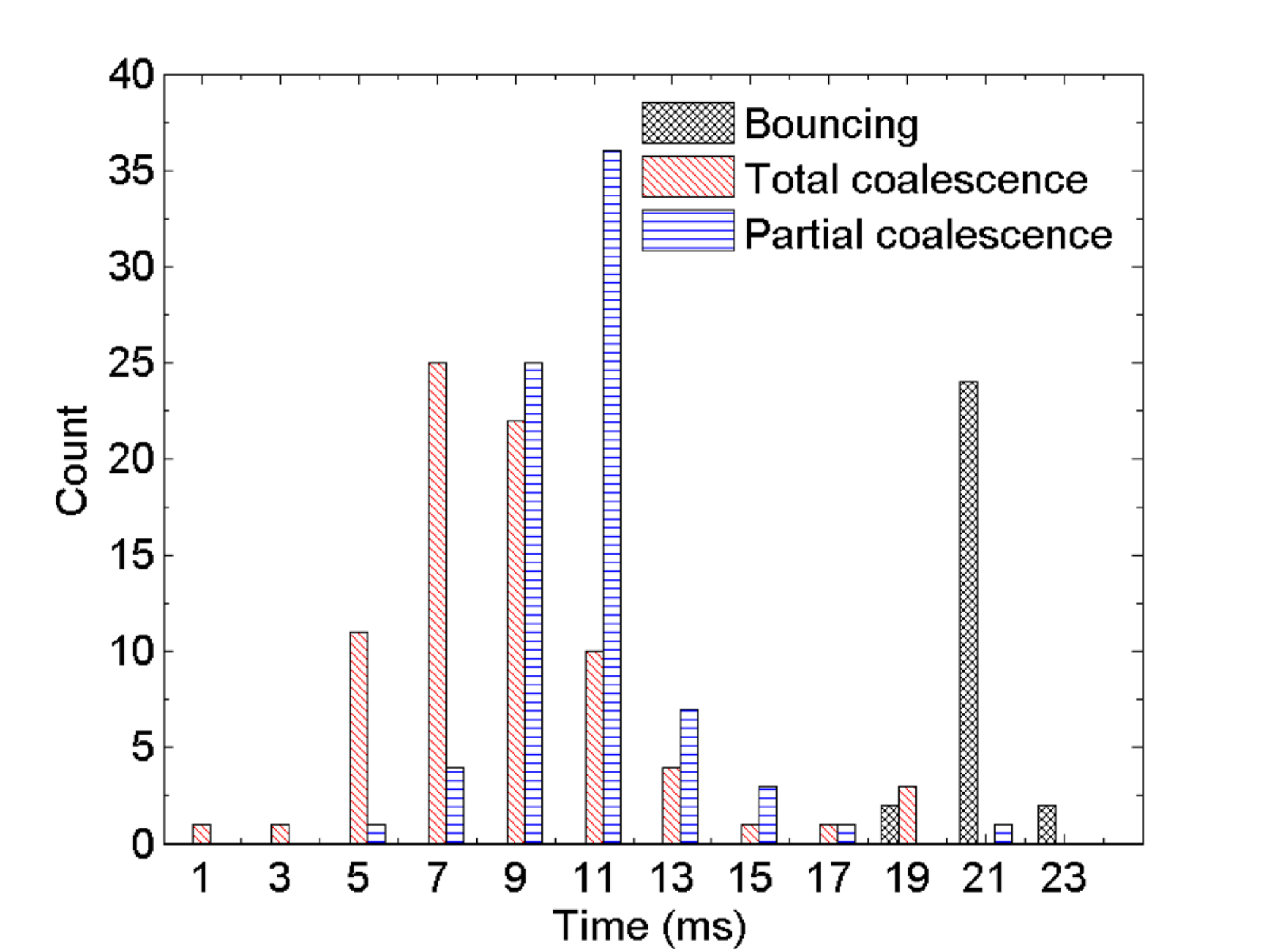}
  \caption{Histogram of the lifetime of the gas layer for three impact phenomena: bouncing, partial coalescence and total coalescence.   The dimensionless parameters are $\text{Oh}=0.0021$, $\text{We}=4$, and $\text{Re}$ in the range of 54--636.   The droplet speed is 0.30 m/s, the flow rate of the falling film is in the range of 0.9--10.7 l/min, and the droplet diameter is 3.3 mm.
  }\label{fig:fig11}
\end{figure}

To further understand the collapse of the gas layer, the lifetime of the gas layer was studied, and a histogram of this lifetime is plotted in Figure \ref{fig:fig11}.  The starting time was determined from the high-speed images, which show noticeable capillary waves on the surfaces of the liquid film upon apparent impact. The ending time was determined from the images which show the rupture of the gas layer indicated by the capillary waves on the droplet and on the liquid film, as explained in Figure \ref{fig:fig06}.
For the bouncing phenomenon, the lift-off of the droplet away from the film was considered as the ending time. Due to this distinctive feature and the fast propagation of the capillary waves in the high-speed images, the life time of the gas layer can be determined at an accuracy of about 0.2--0.4 ms at the filming frame rate 5000 fps. As shown in Figure \ref{fig:fig11}, the bouncing and total coalescence phenomena correspond to long and short gas layer lifetimes, respectively.  This indicates that delaying (accelerating) the rupture of the gas layer results in bouncing (total coalescence).  This is because the speed of the liquid at the top region of the droplet can be retarded via droplet deformation during the thinning of the gas layer.  Following rapid rupture of the gas layer, the fluid near the droplet apex has a higher chance of catching up with the fluid near the bottom of the droplet, preventing the formation of secondary droplets, and leading to total coalescence.

\subsection{Dimple formation and droplet ejection for large droplets}\label{sec:sec3_7}
\begin{figure}
  \centering
  \includegraphics[width=\columnwidth]{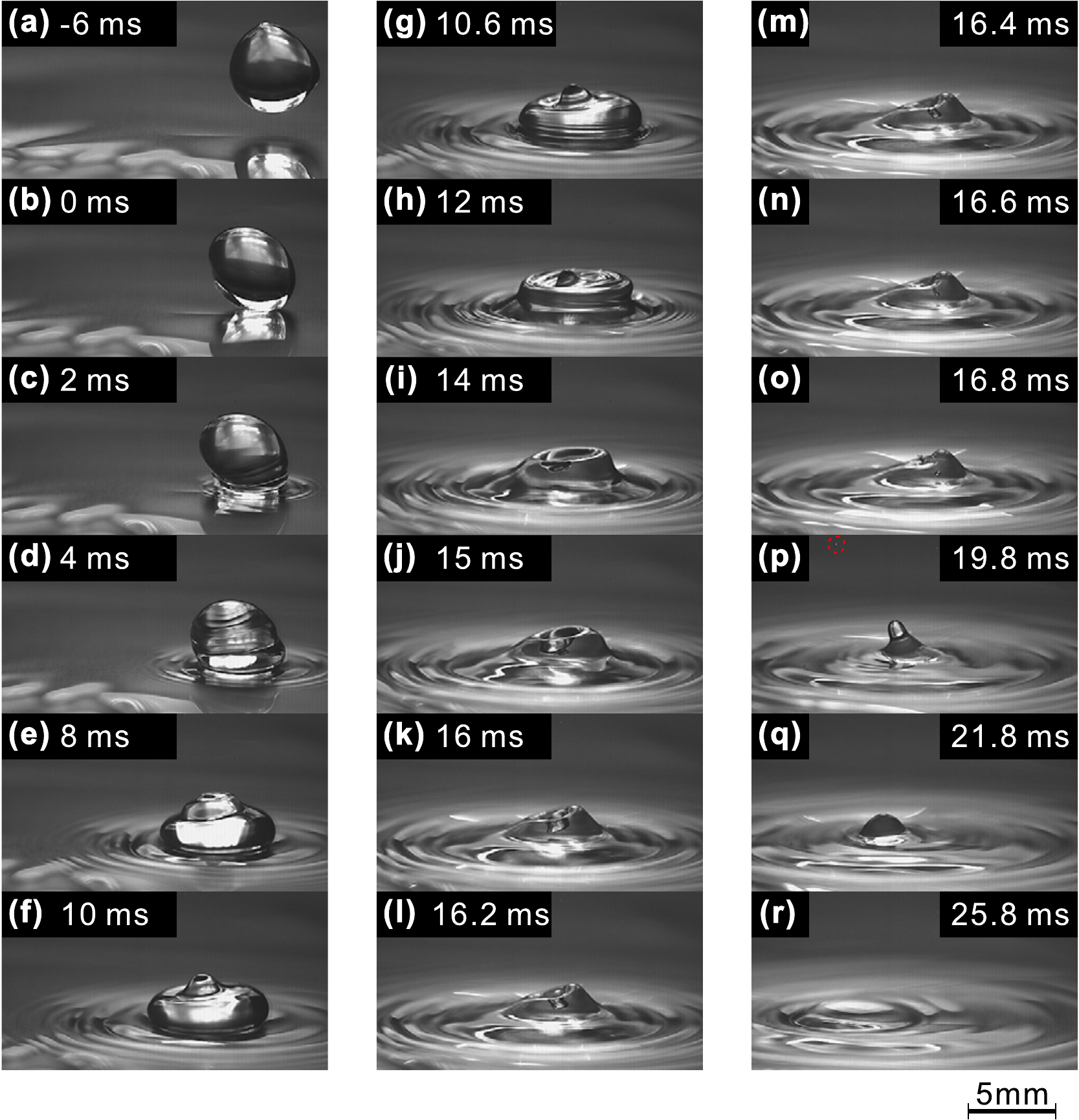}
  \caption{Dimple formation and droplet ejection during the impact of a large droplet on a falling film. The dimensionless parameters are $\text{Oh}=0.0016$, $\text{We}=6.5$, and $\text{Re}=54$. The droplet speed is 0.3 m/s, the flow rate of the falling film is 0.9 l/min, and the droplet diameter is 5.2 mm. See Ref.\ \cite{RefSM}, Movie 5.
  }\label{fig:fig12}
\end{figure}

 The Ohnesorge number can be varied by changing the droplet size.  We consider a case shown in Figure \ref{fig:fig12} characterised by $\text{Oh} = 0.0016$, $\text{We} = 6.5$, and $\text{Re} = 54$. These parameters are similar to those that were used to generate Figure \ref{fig:fig05}, which depicted a case of partial coalescence; the decrease in $\text{Oh}$ and the slight increase in $\text{We}$ were brought about by the increase in the droplet diameter from 3.3 mm to 5.2 mm. The case presented in Figure \ref{fig:fig11} illustrates an impact accompanied by the droplet surface switching from a convex to a concave shape, followed by the ejection of a small droplet. Inspection of Figure \ref{fig:fig11}a shows that the droplet shape prior to impact is far from spherical, as expected for large droplet size, i.e., large $\text{Oh}$ numbers. Upon apparent contact with the liquid film (see Figure \ref{fig:fig12}b), capillary waves are formed and propagate along the surface of the droplet towards the top of the droplet and also on the surface of the liquid film (see Figure \ref{fig:fig12}c-d).  The liquid in the top region of the droplet is less affected by the impact than that near the bottom of the droplet and continues to move towards the film by inertia (see Figure \ref{fig:fig12}e-f).  Following rupture of the gas layer, capillary waves form and propagate on the surface of the droplet towards its apex.  The interaction of these waves with the downwards-moving fluid from the apex leads to the formation of a hat-shaped droplet (see Figure \ref{fig:fig12}g-h) followed by a dimple at the droplet apex (see Figure \ref{fig:fig12}i-k).  The formation of the dimple is facilitated by the relative weakness of capillary forces for such large droplets and their consequent inability to act as effective restoring forces to counterbalance severe interfacial deformation. For the concave interface associated with the dimple, capillarity pulls the liquid at the rim of the dimple causing it to shrink (see Figure \ref{fig:fig12}l-n).  The collision of the interface at the rim of the dimple results in the ejection of a small droplet (see Figure \ref{fig:fig12}o-p).  Due to the flow in the liquid film, the droplet is not symmetric, and the small droplet is ejected in a direction which is not perpendicular to the substrate but deviated in the streamwise direction. The liquid eventually settles down as the energy from the impact is dissipated gradually (see Figure \ref{fig:fig12}q-r).

\subsection{The asymmetric splashing process}\label{sec:sec3_8}
\begin{figure}
  \centering
  \includegraphics[width=\columnwidth]{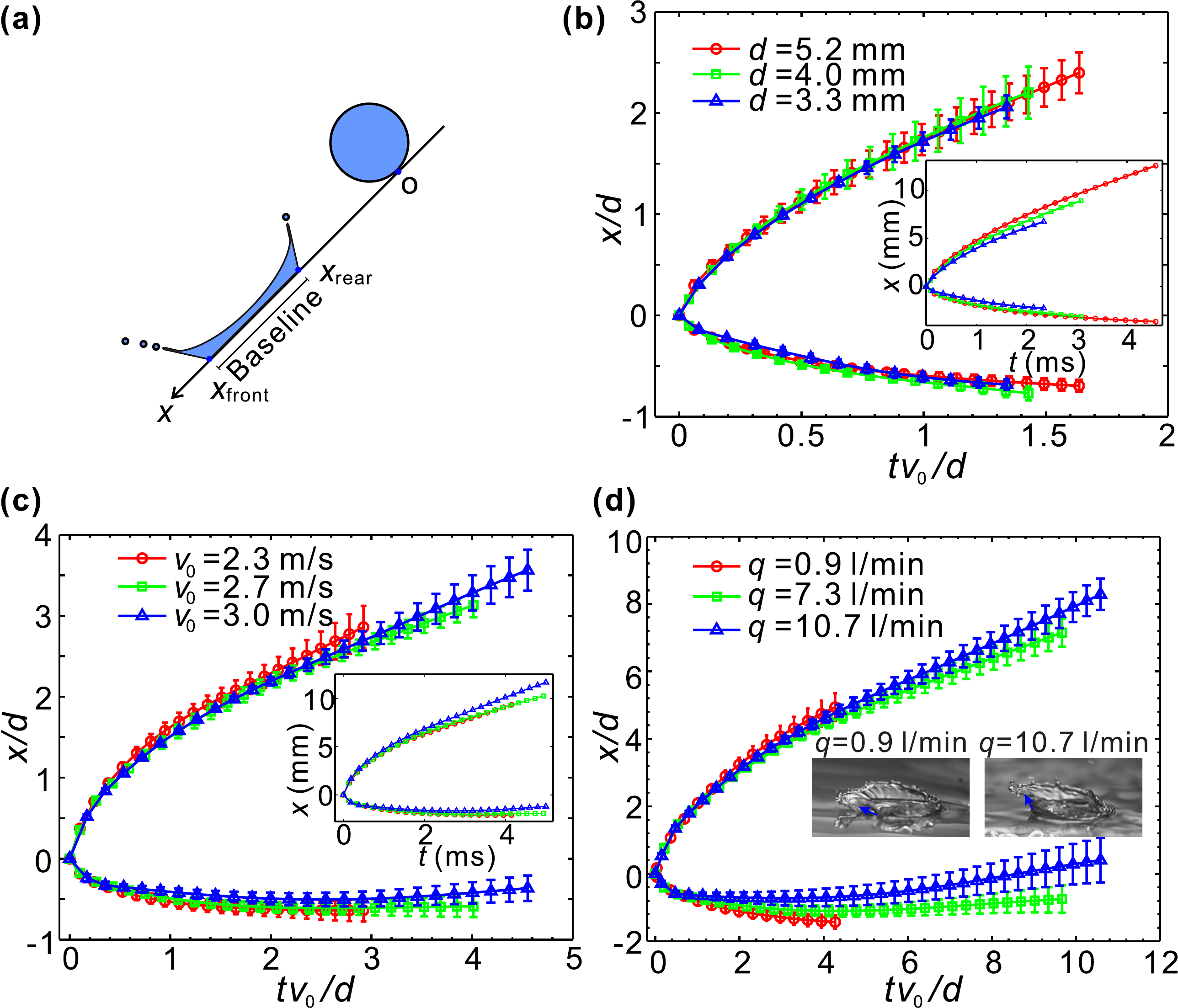}
  \caption{
  Effects of the droplet size, the droplet speed, and the film flow rate on the propagation of the baseline during droplet splashing.  (a) Schematic diagram of the definition of the coordinate system.  (b) Effect of the droplet size. The droplet speed is 1.87 m/s, the film flow rate is 0.9 l/min, and correspondingly,  $\text{Re}=54$ and $\text{We}$ in the range of 156--254. (c) Effect of the droplet speed. The film flow rate is 4.1 l/min, the droplet diameter is 3.3 mm, and correspondingly, $\text{Re}=245$ and $\text{Oh}=0.0021$. (d) Effect of the film flow rate. The droplet speed is 2.67 m/s, and the droplet radius is 3.3 mm, and correspondingly, $\text{Oh}=0.0021$ and $\text{We}= 316$. The inset shows two images of crowns at different film flow rates.
  }\label{fig:fig13}
\end{figure}

We devote the concluding part of this section to quantitatively examining the details of
asymmetric
splashing at different film flow rates, droplet speeds, and droplet sizes. The asymmetric splashing process was studied to consider the effects of relevant parameters and quantified by the propagation of the baseline of the droplet, as shown in Figure \ref{fig:fig13}a. The streamwise direction of the falling film is defined as the positive direction of the movement of the baseline, and the initial contact point of the droplet on the falling liquid film is defined as the origin.  The initial contact time is determined by interpolating the interface position from the image sequence, which provides a higher precision of the initial contact time than the time delay between two consecutive frames. The length of the baseline of the droplet, and the position of the front and rear point is normalized by the diameter of the droplet, and the time is normalized by the speed and the diameter of the droplet:
\begin{equation}\label{eq:x}
  \hat{x}=x/d
\end{equation}
\begin{equation}\label{eq:t}
  \hat{t}=t{{v}_{0}}/d
\end{equation}
Since the droplet impact in the present study is different from the perpendicular impact of droplets on quiescent films, the propagation of the front and rear of the baselines are asymmetric, as shown in Figure \ref{fig:fig13}. The degree of asymmetry is influenced by the oblique impact of the droplet and the flow rate of the liquid film. Upon impact, the front point of the baseline moves along the flow direction, and the rear point of the baseline moves backwards. In addition, the front point moves at a higher speed than the rear point.

The effect of the droplet size on the asymmetric splashing process is shown in Figure \ref{fig:fig13}b.  The curves of the baselines for different droplet sizes collapse, even though the dimensional data show clear difference at different droplet sizes (see the inset plot of Figure \ref{fig:fig13}b).  For smaller droplets, the baselines propagate much more slowly at both the front and the rear of the droplets, and the splashing process is much shorter than that for large droplets.  This is because the mass and the momentum introduced into the film by small droplets are much less than those by large droplets.

The propagation of the baseline for droplet impact at different droplet speeds is shown in Figure \ref{fig:fig13}c. The propagation of the baseline appears to be weakly dependent on the droplet speed. The dimensional plot (see the inset plot of Figure \ref{fig:fig13}c) shows that with increasing the droplet speed, both the front and the rear points move faster in the downstream direction.  This is mainly due to the fact that droplets with higher speeds (larger momentum) can overcome the resistance of impact and move faster downstream than droplets with lower speeds.

The effect of the film flow rate on the propagation of the baseline is shown in Figure \ref{fig:fig13}d.  At a small flow rate, the liquid film is thin, and the liquid sheet of the crown is ejected close to the substrate.  In contrast, at a high flow rate, the liquid film is thicker, and the liquid sheet of the crown is ejected close to the direction perpendicular to the substrate, as shown in the inset images of Figure \ref{fig:fig13}d. Similar phenomena also exist for droplet impact on quiescent liquid films \cite{Wang2000thinLiquidFilm}. Therefore, at a low film flow rate, the movement of the liquid sheet of the crown can promote the movement of the droplet baseline.  With increasing the film flow rate, the film thickness increases, and the motion enhancing effect of the crown decrease.  Consequently, the speeds of the front and rear points reduce.  If the film flow rate increases further, the flow speed in the falling film also increases to the point where the film simply transports the droplet downstream so that both the front and rear points move in the streamwise direction.

\section{Conclusion}\label{sec:sec4}
In this paper, the oblique impact of droplets on inclined falling liquid films was studied using high-speed imaging.  The results of droplet impact on flowing liquid films showed unique features different from the impact process on quiescent liquids.  Under different impact conditions, various impact phenomena were observed: droplet bouncing, partial coalescence, total coalescence, and splashing.  An impact regime map was generated for the impact of droplets on falling liquid films.  Splashing phenomena occur at large Weber numbers, $\text{We}$, while other phenomena manifest themselves at low $\text{We}$. Low Reynolds numbers of the liquid films, $\text{Re}$, are associated with partial coalescence, while the probability of bouncing and total coalescence increases with  $\text{Re}$. Dimple formation and droplet ejection were observed for the impact of large droplets. Splashing processes were quantified through the propagation of the droplet baselines during impact, and the effects of the droplet size, the droplet speed, and the flow rate of the falling liquid film were analysed.
The analysis of the lubrication force during the impact process shows that a higher flow rate in the liquid film produces a larger lubrication force, slows down the drainage process, and increases the probability of droplet bouncing.

The study mainly focused on the fluid dynamics of the droplet impact on falling liquid films.  There are many open questions related to droplet impact that are worthy of investigation, such as the direct measurement of the flow field during the impact process, the evolution of the gas layer, and the effects of droplet impact on the flow in the liquid films. In various relevant applications, such as annular flow, spray cooling, and spray painting, the impact processes are much more complex. For example, the impact of swarms of droplets on liquid films involves the interaction of numerous impact processes; the flow of the gas phase may induce turbulence in the liquid films and droplet entrainment on the film surface.  This investigation will deepen our understanding of the physics involved in the phenomena of droplet impact, leading to a marked improvement in processes and applications that rely on these phenomena.

\section*{Acknowledgements}
We would like to acknowledge the support of the Engineering and Physical Sciences Research Council, UK, through the Programme Grant, MEMPHIS (EP/K003976/1).

\bibliography{DropletImpact}
\end{document}